\documentclass{article}
\usepackage{graphicx}
\usepackage{amssymb}
\usepackage{amsmath}
\input epsf.tex   

\def\beq{\begin{equation}} 
\def\eeq{\end{equation}} 
\def\bea{\begin{eqnarray}} 
\def\eea{\end{eqnarray}}

\newcommand{\gsim}{\stackrel{>}{_\sim}}
\newcommand{\lsim}{\stackrel{<}{_\sim}}
\def\openone{\leavevmode\hbox{\small1\kern-7.3pt\normalsize1}}%

\def\abstracts#1{
\begin{center}
\vskip.6in
ABSTRACT
\vskip.3in
{\begin{minipage}{4.7truein}
                 \parindent=0pt #1\par
                 \end{minipage}}\end{center}
                 \vskip 2em \par}

\begin{document}

\title{Little Higgs Review}

\date{February 18, 2005}

\author{Martin Schmaltz $^{a}$\thanks{
Supported by DOE-OJI DE-FG02-91ER40676 and
Alfred P. Sloan Research Fellowship.} \quad
and \ David Tucker-Smith $^{b}$\thanks{Work
supported by a Research Corporation
Cottrell College Science Award.}}

\maketitle

\begin{center}{$^a$ Physics Department, Boston University \\
Boston, MA  02215, USA \\
{\tt schmaltz@bu.edu}
\vskip.1in
$^b$ Department of Physics, Williams College \\ 
Williamstown, MA  01267, USA \\
{\tt dtuckers@williams.edu}}
\end{center}

\abstracts{
Recently there has been renewed interest in the possibility that the Higgs
particle of the Standard Model is a pseudo-Nambu-Goldstone boson.  This
development was spurred by the observation that if certain global symmetries
are broken only by the interplay between two or more coupling constants, then
the Higgs mass-squared is free from quadratic divergences at one loop.  This
``collective'' symmetry breaking  is the essential ingredient  in little
Higgs theories,  which are weakly coupled extensions of the Standard Model
with little or no fine tuning, describing physics up to an energy scale $\sim
10~{\rm TeV}$.  Here we give a pedagogical introduction to little Higgs
theories.  We review their structure and phenomenology, focusing mainly on
the $SU(3)$ theory, the Minimal Moose, and the Littlest Higgs as concrete examples.
}

\newpage

\tableofcontents

\section{INTRODUCTION} 

A few years before the start of the LHC program, electroweak symmetry breaking
remains poorly understood. The detailed quantitative fit of Standard Model
predictions
to precision electroweak data strongly suggests
that electroweak symmetry
is broken by one or more weakly coupled Higgs doublets.
However, fundamental scalar particles suffer from a radiative
instability in their masses,
leading us to expect additional structure
(such as compositeness, supersymmetry, little Higgs, ...)
near the weak scale. 

Interestingly, we can turn this problem into a prediction for the LHC.
The argument goes as follows:
let us assume that precision electroweak data are indeed telling
us that there are no new particles beyond the Standard Model
(with the exception of possible additional Higgs doublets) with
masses at or below the weak scale. Then physics at the weak scale
may be described by an ``effective Standard Model" which has the
particle content of the Standard Model and in which possible new
physics is parametrized by higher-dimensional operators suppressed by
the new physics scale $\Lambda\gsim $ TeV.
All renormalizable couplings are as in the Standard Model.
If there are additional Higgs fields then more complicated
Higgs self-couplings as well as Yukawa couplings are possible.
Since no Higgs particles have been discovered so far, the effects
of additional Higgs fields can be parametrized by effective
operators for the Standard Model fields. 

The higher-dimensional operators can be categorized by
the symmetries which they
break. The relevant symmetries are baryon and lepton
number ($B$ and $L$), CP and
flavor symmetries, and custodial $SU(2)$ symmetry.
The wealth of indirect experimental
data can then be translated into bounds on the
scale suppressing the operators
\cite{Eidelman:2004wy,Group:2004qh,D'Ambrosio:2002ex,Han:2004az}.
Examples of such operators and the resulting bounds are summarized
in Table \ref{tab:operatorbounds}.
The bounds imply that physics at the TeV scale has to
conserve B and L, flavor and CP to a very high accuracy, and that
violations of custodial symmetry and contributions to the 
S-parameter should also be small.

\begin{table}[ph]
\vskip.1in
\centering{\footnotesize
\begin{tabular}{@{}lll@{}}
\hline
{} &{} &{}\\[-1.5ex]
{broken symmetry} & operators & scale $\Lambda$\\[1ex]
\hline
{} &{} &{}\\[-1.5ex]
B, L& $(QQQL) / \Lambda^2$ & $10^{13}$ TeV \\[1ex]
flavor (1,2$^{\rm nd}$ family), CP &
$(\bar d s \bar d s)/ \Lambda^2$ & 1000 TeV \\[1ex]
flavor (2,3$^{\rm rd}$ family) &
$m_b (\bar s \sigma_{\mu \nu} F^{\mu \nu} b)/ \Lambda^2$ & 50 TeV \\[1ex]
custodial $SU(2)$ & $(h^\dagger D_\mu h)^2 / \Lambda^2$ & 5 TeV \\[1ex]
none (S-parameter)& $(D^2 h^\dagger D^2 h) /  \Lambda^2$ & 5 TeV\\[1ex]
\hline
\end{tabular}\label{tab:operatorbounds} }
\caption{Lower bounds on the scale which suppresses higher-dimensional
operators that violate approximate symmetries of the Standard Model.}
\vspace*{-1pt}
\end{table}

\begin{figure}[htb]
\centerline{
{\includegraphics[width=3.8in]{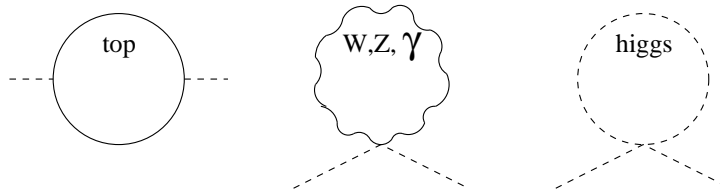}}
}
\caption{ The most significant quadratically divergent contributions 
to the Higgs mass in the Standard Model. }
\label{fig:smloops}
\end{figure}

The question then becomes
if it is possible to add new physics at the TeV scale to the
SM which stabilizes the Higgs mass but does not violate the
above bounds. To understand the requirements on this new physics
better we must look at the source of the Higgs mass instability.
The three most dangerous radiative corrections to the Higgs mass
in the Standard Model come from one-loop diagrams with top quarks,
$SU(2)$ gauge bosons, and the Higgs itself
running in the loop (Figure \ref{fig:smloops}).

All other diagrams give smaller contributions because they
involve small coupling constants. 
Assuming that the Standard Model remains valid up to a cut-off  of order the LHC center-of-mass energy, $\Lambda \sim 10~{\rm TeV}$, the
three diagrams give
\bea
{\rm top\ loop} &\quad
-\frac{3}{8 \pi^2} \lambda_t^2 \Lambda^2\ &\sim\ -(2\ {\rm TeV})^2 
\nonumber \\
SU(2) {\rm\ gauge\ boson\ loops} &\quad
{9 \over 64 \pi^2} g^2 \Lambda^2\ &\sim \ (700\ {\rm GeV})^2
\nonumber \\
{\rm Higgs \ loop} & \quad 
{1 \over 16 \pi^2} \lambda^2 \Lambda^2\ &\sim \ (500\ {\rm GeV})^2.
\nonumber
\eea
The total Higgs mass-squared includes the sum of these loop
contributions and a tree-level mass-squared parameter.

To obtain a weak-scale  expectation value for the Higgs  
without worse than 10\% fine tuning, the top, gauge, and Higgs loops must be cut off at scales satisfying
\bea
\Lambda_{top} \lsim 2\ {\rm TeV} \qquad \Lambda_{gauge} \lsim 5\ {\rm TeV}
\qquad \Lambda_{Higgs} \lsim 10\ {\rm TeV}.
\eea
We see that the Standard Model with a cut-off near the maximum attainable
energy at the Tevatron ($\sim 1$ TeV) is natural, and
we should not be surprised that we have not observed any new physics.
However, the Standard Model with a cut-off of order the LHC energy would be fine
tuned, and so we should expect to see new physics at the
LHC.

More specifically, we expect new physics that cuts off the divergent
top loop at or below 2 TeV. In a weakly coupled theory
this implies that there are new particles
with masses at or below 2 TeV. These particles must couple to the Higgs,
giving rise to a new loop diagram that cancels the quadratically divergent 
contribution from the top loop. For this cancellation to be
natural, the new particles 
must be related to the top quark by some symmetry, implying that the new particles have similar quantum numbers
to top quarks. Thus naturalness arguments predict a new multiplet
of colored particles with mass below 2 TeV, particles that would be
easily produced at the LHC. In supersymmetry
these new particles are of course the top squarks.

Similarly, the contributions from $SU(2)$ gauge loops must be canceled
by new particles related to the Standard Model $SU(2)$
gauge bosons by symmetry, and the masses of these particles must
be at or below 5 TeV for the cancellation to be natural.
Finally, the Higgs loop requires new particles related to the Higgs itself
at or below 10 TeV. 
Given the LHC's 14 TeV center-of-mass energy, 
these predictions are very exciting,
and encourage us to explore different possibilities
for what the new particles could be.

It could be that the new particles are the superpartners
predicted by the Minimal Supersymmetric
Standard Model \cite{mssm}. There the quadratic divergence of the top
quark loop is canceled by a corresponding loop
with top squarks:  supersymmetry predicts the necessary relationship
between the  top and stop coupling constants. 
Or, it could be that the Higgs is a composite resonance
at the TeV scale as in technicolor \cite{technicolor} or
composite Higgs models \cite{Kaplan:1983fs,Kaplan:1983sm,Georgi:1984af}.
Or perhaps extra dimensions are
lurking at the TeV scale, with new mechanisms to stabilize the Higgs mass \cite{extradimensions}.

Here we explore a different possibility, that the Higgs is a
pseudo-Nambu-Goldstone boson.  This idea was first  suggested in \cite{Georgi:1974yw,Georgi:1975tz}, and was recently revived by Arkani-Hamed, Cohen and Georgi
when they constructed the first successful
``little Higgs'' model \cite{ACG}, and thereby started an industry
of ``little model building'' \cite{Arkani-Hamed:2002pa, Arkani-Hamed:2002qx, Arkani-Hamed:2002qy,  Gregoire:2002ra, Low:2002ws,  amsterdam, kaplanschmaltz, Chang:2003un, skibaterning, Chang:2003zn, Contino:2003ve, Cheng:2003ju, Katz:2003sn, Birkedal:2004xi, Cheng:2004yc, Kaplan:2004cr, simplest, Low:2004xc,  Agashe:2004rs, Batra:2004ah, Thaler:2005en}.
As we will see, in these theories the Higgs mass is protected from one-loop quadratic divergences by  approximate global symmetries under which the Higgs field shifts.  New particles must be introduced to ensure that the global symmetries are not broken too severely, and these are the states  that cut off the quadratically divergent top, gauge, and Higgs loops.

The outline of the rest of the article is as follows:  in the next section we review some basic aspects of Nambu-Goldstone-Bosons. Then, we describe in section 3 how to construct a little Higgs theory, using the $SU(3)$ model of Refs.~\cite{amsterdam,kaplanschmaltz,simplest} as an illustrative example.  In section 4 we discuss the prototype product-group models of Refs.~\cite{Arkani-Hamed:2002qx, Arkani-Hamed:2002qy}, along with several variations.  Finally, in section 5 we discuss precision electroweak constraints on little Higgs theories and the prospects for discovering the new particles  they predict at the LHC.

\section{NAMBU-GOLDSTONE BOSONS}

Nambu-Goldstone bosons (NGBs) arise whenever a continuous global
symmetry is spontaneously broken.
If the symmetry is exact, the NGBs are exactly massless and have
only derivative couplings.

\begin{figure}[htb]
\vskip 0.2truein
\centerline{
{\includegraphics[width=2in]{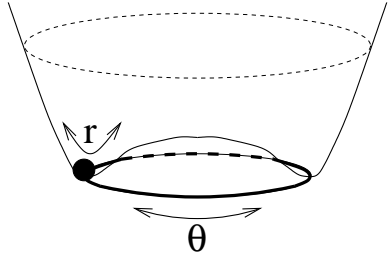}}}
\vskip- 0.1truein
\caption{ The ``Mexican hat'' potential for $\Phi$. The
black dot represents the vacuum expectation value $f$, $r$ is the radial
mode and $\theta$ the Nambu-Goldstone boson. }
\label{fig:mexhat}
\end{figure}

\vskip.2in\noindent
\underline{$U(1)$ example:} Consider for example
a theory with a single complex scalar field $\phi$ with potential
$V=V(\phi^* \phi)$. The kinetic 
energy term $\partial_\mu \phi^* \partial^\mu \phi$ and the potential are invariant
under the $U(1)$ symmetry transformation
\bea
\phi \rightarrow e^{i \alpha} \phi.
\label{eq:u1trafo}
\eea
If the minimum of the potential is not at the origin but at
some distance $f$ away as in the 
famous ``wine bottle" or ``Mexican hat" potential (Figure \ref{fig:mexhat}),
then the $U(1)$ symmetry is spontaneously broken in the vacuum.
When we expand the field for small fluctuations around its 
vacuum expectation value (VEV), it takes the form
\bea
\phi(x)=\frac{1}{\sqrt{2}} (f+r(x))\ e^{i \theta(x)/f},
\eea
where $f$ is the VEV, $r(x)$ is the massive ``radial mode" and
$\theta(x)$ is the NGB. The factor of $1/\sqrt{2}$ ensures canonical kinetic
terms for the real fields $r$ and $\theta$.

The radial field $r$ is invariant under the $U(1)$ symmetry transformation of
Eq.~(\ref{eq:u1trafo}), whereas the NGB field $\theta$ shifts,
\bea
\theta \rightarrow \theta +\alpha
\eea
under $U(1)$ transformations. We say that the $U(1)$ symmetry is
non-linearly realized. Suppose that we now integrate out the massive field $r$.  We can be sure that the  resulting effective Lagrangian for the NGB $\theta(x)$
will not include a mass term for $\theta$, or any potential terms for that matter, because 
the shift symmetry forbids all non-derivative couplings
of $\theta$.

\vskip.2in
\noindent
\underline{Non-Abelian examples:}  Generalizing to
spontaneously broken non-Abelian symmetries, we find one NGB for every broken
symmetry generator. For example,
suppose we break $SU(N) \rightarrow SU(N-1)$ with the VEV of a 
single fundamental $\phi$ of $SU(N)$.
The number of broken generators is the total number of
generators of $SU(N)$ minus the number
of unbroken generators, i.e.
\bea
[N^2-1]-[(N-1)^2-1]=2N-1.
\eea
The NGBs are conveniently parametrized by writing
\bea
\phi=exp\left\{{i\over f}
\left(\begin{array}{c|c} 
& \pi_1\\ & \vdots \\ & \pi_{N-1} \\ \hline \pi^*_1 \cdots \pi_{N-1}^* & \pi_0/\sqrt{2}
\end{array}\right)
\right\}
\left(\begin{array}{c} 0 \\ \vdots \\ 0 \\ f  \end{array}\right)
\equiv e^{i \pi/f} \phi_0.
\label{eq:ngbparam}
\eea
The field $\pi_0$ is real whereas the the fields
$\pi_1 \cdots \pi_{N-1}$ are complex. The last equality defines a convenient
short-hand notation which we will employ whenever the precise form of $\pi$ and 
$\phi_0$ is clear from the context.

Another example of symmetry breaking and NGBs which has found 
applications in little Higgs model building is
\bea
SU(N)\rightarrow SO(N) \ .
\eea
Here the number of NGBs is the number of fields in the adjoint of $SU(N)$ minus the
number of fields in the adjoint of $SO(N)$ (antisymmetric tensor), i.e.
\bea
[N^2-1]-{N(N-1)\over 2} = {N(N+1)\over 2} -1 \ .
\eea
For even $N$ we also have
\bea
SU(N)\rightarrow SP(N). 
\eea
and the number of NGBs is the number of fields in the adjoint of $SU(N)$ minus the
number of fields in the adjoint of $SP(N)$ (symmetric tensor), i.e.
\bea
[N^2-1]-{N(N+1)\over 2} = {N(N-1)\over 2} -1\ .
\eea
Finally, for
\bea
SU(N)\times SU(N)\rightarrow SU(N),
\eea
the number of NGBs is
\bea
2[N^2-1] - [N^2-1]=N^2-1 \ .
\eea
In this last case the symmetry breaking can be achieved by a
VEV which transforms as a bi-fundamental
under the two $SU(N)$ symmetries. Denoting transformation
matrices of the two $SU(N)$ as
$L$ and $R$ respectively we have
\bea
\phi \rightarrow L\, \phi\, R^\dagger \ .
\eea
The symmetry-breaking VEV is proportional to the unit matrix
\bea
<\! \phi \!> \equiv \phi_0 =
f\ \left(\begin{array}{ccc} 1& &0 \\ & \ddots & \\ 0 && 1  \end{array}\right).
\eea
This VEV is left invariant under ``vector" transformations for which $L=R\equiv U$,
\bea
\phi_0\ \longrightarrow U\, \phi_0 \ U^\dagger = \phi_0 \ ,
\eea
while other symmetry generators (the ``axial" generators) are broken.  The corresponding
NGBs can be parametrized as
\bea
\phi = \phi_0\, e^{i \pi/f}= f\, e^{i \pi/f},
\eea
where $\pi$ is a Hermitian traceless matrix with $N^2-1$ independent components.

\subsection{How do NGBs transform ?}

We now show how NGBs transform under the broken and unbroken symmetries
in the example of $SU(N)\rightarrow SU(N-1)$, which is often denoted in more
mathematical notation as $SU(N)/SU(N-1)$.
The NGBs can be parametrized as $\phi \equiv e^{i \pi} \phi_0$ as in Eq.~(\ref{eq:ngbparam}).
Let's consider first the unbroken $SU(N-1)$ transformations. Under these transformations we have
\bea 
\phi \rightarrow U_{N-1}\ \phi = 
(U_{N-1}\ e^{i \pi}\ U_{N-1}^\dagger)\, U_{N-1}\ \phi_0 =
e^{i (U_{N-1}  \,\pi\, U_{N-1}^\dagger)}\, \phi_0, 
\eea
where in the second equality we used the fact that the symmetry-breaking $\phi_0$ is
invariant under the unbroken $U_{N-1}$ transformations. Therefore
the NGBs transform in the usual ``linear" way under $SU(N-1)$,
$\pi \rightarrow U_{N-1} \, \pi\, U_{N-1}^\dagger$. Explicitly, 
the unbroken $SU(N-1)$ transformations are
\bea
U_{N-1} =
\left(\begin{array}{cc} \hat U_{N-1} & 0 \\  0 & 1
\end{array}\right) \ .
\eea
The single real NGB $\pi_0$ transforms as a singlet, whereas the
$N-1$ complex NGBs transform as
\bea
\left(\begin{array}{cc|c} 0& & \vec \pi \\ &&\\ \hline \vec \pi^\dagger& & 0 \end{array}\right)
\rightarrow 
U_{N\!-\!1}
\left(\begin{array}{cc|c} 0& & \vec \pi \\ &&\\ \hline \vec \pi^\dagger& & 0 \end{array}\right)
U_{N\!-\!1}^\dagger =
\left(\begin{array}{cc|c} 0& & \hat U_{N\!-\!1} \vec \pi \\ 
&&\\ \hline \vec \pi^\dagger \hat U_{N\!-\!1}^\dagger& & 0 \end{array}
\right)
\eea
where we have used a vector notation $\vec \pi$ to represent the
$N\!-\!1$ complex NGBs as a column vector.
We see that $\vec \pi \rightarrow \hat U_{N-1} \vec \pi$, i.e. $\vec \pi$
transforms in the fundamental representation of $SU(N\!-\!1)$. 

Under the broken symmetry transformations we have
\bea
\phi \rightarrow U \, e^{i \pi} \, \phi_0 &=& exp \left\{i 
\left(\begin{array}{cc} 0 & \vec \alpha \\ \vec \alpha^\dagger & 0 \end{array}\right)
\right\}
exp\left\{i 
\left(\begin{array}{cc} 0 & \vec \pi \\ \vec \pi^\dagger & 0 \end{array}\right)
\right\}
\phi_0  \nonumber \\
&\equiv&
exp\left\{i 
\left(\begin{array}{cc} 0 & {\vec \pi'} \\ {{\vec \pi'}}\,\!^\dagger & 0 \end{array}\right)
\right\} 
U_{N-1}(\vec\alpha, \vec\pi) \ 
\phi_0  \nonumber \\
&=&
exp\left\{i 
\left(\begin{array}{cc} 0 & {\vec \pi'} \\ {{\vec \pi}'}\,\!^\dagger & 0 \end{array}\right)
\right\} 
\phi_0,
\label{eq:brokentrafo}
\eea
where in the second equality we used the fact that any $SU(N)$
transformation can be written
as the product of a transformation in the coset $SU(N)/SU(N-1)$
times an $SU(N-1)$ transformation
\cite{CCWZ}.
The $U_{N-1}(\vec \alpha, \vec \pi)$ transformation, which
depends on $\vec \alpha$ and $\vec \pi$,
leaves $\phi_0$ invariant and can therefore be removed.
Equation~(\ref{eq:brokentrafo}) defines the transformed field
$\vec \pi'$, and in general, $\vec \pi'$ is a complicated function
of $\vec \alpha$ and $\vec \pi$.
To linear order the transformation is simple,
\bea
\vec \pi \rightarrow \vec \pi' = \vec \pi + \vec \alpha,
\eea
showing that the NGBs shift under the non-linearly
realized symmetry transformations.
As in the $U(1)$ case, the shift symmetry ensures that
NGBs can only have derivative interactions.

\subsection{Effective Lagrangian for NGBs}

Our goal for this section is to write the most general effective
Lagrangian involving only the massless NGB fields, which respects the full
$SU(N)$ symmetry. This is where the utility of the exponentiated
fields $\phi$ becomes obvious: while
the full $SU(N)$ transformations on the $\pi$'s are complicated,  
the $\phi$'s transform very simply. To get the low energy effective Lagrangian
we expand in powers of $\partial_\mu/ \Lambda$ and
write the most general possible $SU(N)$-invariant function of
$\phi=e^{i \pi/f} \phi_0$  at every order.
With no derivatives we can form two basic
gauge invariant objects, $\phi^\dagger \phi=f^2$ and
$\epsilon^{a_1 ...\,a_N} \phi_{a_1} \phi_{a_2} \cdots \phi_{a_N} = 0$.
Thus the most general invariant contribution to the potential is simply
a constant. 
You can convince yourself that the most general term that can be written
at quadratic order is a constant times $|\partial_\mu \phi|^2$ and
therefore, we have
\bea
\mathcal L =  {\rm const.} + f^2 |\partial_\mu \phi|^2 + \mathcal{O}(\partial^4),
\eea
where we normalized the coefficient of the second-order term such
that the $\pi$ fields have canonical kinetic terms. The
kinetic term for $\phi$ expanded to higher-order in the $\pi$ fields
contains interactions that determine the scattering of arbitrary numbers
of $\pi$'s at low energies, in terms of the single parameter $f$.

\section{CONSTRUCTING A LITTLE HIGGS: SU(3)}

Now that we know how to write a Lagrangian for NGBs we would like to
use this knowledge to write a model where the Higgs is a NGB. The
explicit model we are going to construct in this section
is the ``simplest little Higgs'' \cite{amsterdam,kaplanschmaltz,simplest}.
Consider the symmetry breaking
pattern $SU(3)/SU(2)$, with NGBs
\bea
\pi = \left(\begin{array}{c|c}
\begin{array}{cc} -\eta/2 & 0 \\ 0 & -\eta/2 \end{array}
& h \\ \hline h^\dagger & \eta \end{array}\right).
\eea
Note that $h$ is a doublet under the unbroken $SU(2)$, as required
for the Standard Model Higgs, and it is also an NGB --
it shifts under ``broken'' $SU(3)$ transformations. The field
$\eta$ is an $SU(2)$ singlet, which we will ignore for simplicity
in most of the following.
To see what interactions we get for $h$, we expand
\bea
\phi= exp\left\{{i\over f}
\left(\begin{array}{cc}
0& h \\ h^\dagger & 0 \end{array}\right) \right\}
\left(\!\begin{array}{c} 0 \\ f \end{array}\!\right)
=
\left(\!\begin{array}{c} 0 \\ f \end{array}\!\right)
+ {i} \left(\!\begin{array}{c} h \\ 0 \end{array}\!\right)
-{1\over 2 f} \left(\!\begin{array}{c} 0 \\ h^\dagger h \end{array}\!\right)
+ \cdots,
\eea
and therefore obtain
\bea
f^2 |\partial_\mu \phi|^2 = 
|\partial_\mu h|^2 + {|\partial_\mu h|^2 h^\dagger h \over f^2} + \cdots, 
\label{eq:higgslagrangian}
\eea
which contains the Higgs kinetic term as well as interactions suppressed 
by the symmetry-breaking scale $f$. 

Because the Lagrangian contains
non-renormalizable interactions, it can only be an effective low-energy 
description of physics. To determine the cut-off $\Lambda$ at which
the theory becomes strongly coupled, we can compute a loop and ask
at what scale it becomes  as important as a corresponding tree-level 
diagram. The simplest example is the quadratically divergent one-loop
contribution to the kinetic term that stems from contracting
$h^\dagger h$  into a loop in the second term in
Eq.~(\ref{eq:higgslagrangian}). 
Cutting the divergence off at $\Lambda$ we find a renormalization of
the kinetic term proportional to 
\bea
{1 \over f^2} {\Lambda^2 \over 16 \pi^2},
\eea
and therefore $\Lambda \lsim 4\pi f$.

Summarizing, we now have a theory which produces a ``Higgs''
doublet transforming
under an exactly preserved (global)  $SU(2)$. This ``Higgs'' is 
a NGB and therefore exactly massless. It has non-renormalizable
interactions suppressed by the scale $f$, which become strongly
coupled at $\Lambda=4\pi f$. Because of the shift symmetry,
no diagrams, divergent or not, can give rise to a mass for $h$.
Of course, at this point the theory is still very far from what
we want:  an NGB can only have derivative interactions, which means no
gauge interactions, no Yukawa couplings and no quartic potential.
Any of these interactions explicitly break the
shift symmetry $h \rightarrow h + {\rm const}$.
In the following subsections we discuss how to add these
interactions without re-introducing one-loop quadratic divergences.

\subsection{Gauge interactions}

Let us try to introduce the $SU(2)$ gauge interactions for $h$
(we ignore hypercharge for the moment, it will be easy to add later).
To do so we simply follow our nose and see where it leads us.
We will arrive at the right answer after a few unsuccessful
attempts.

\underline{First attempt:} Let's simply couple $h$ to $SU(2)$
gauge bosons in the usual way, by adding to the
Lagrangian of  Eq.~(\ref{eq:higgslagrangian}) the term
\bea
|g W_\mu h|^2,
\label{eq:gaugestupid1}
\eea
along with another term with one derivative and one $SU(2)$ gauge boson $W_\mu$,
as required by gauge invariance.
\begin{figure}[htb]
\vskip 0.2truein
\centerline{\epsfysize=1.3in
{\includegraphics[width=3in]{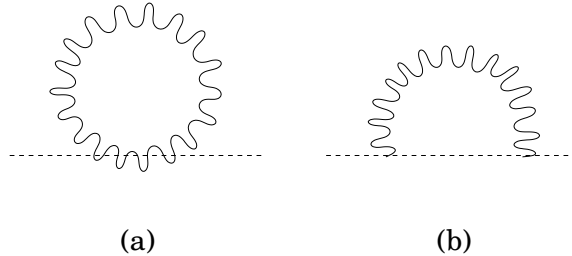}}}
\vskip- 0.1truein
\caption{Quadratically divergent gauge loop contributions
to the Higgs mass. }
\label{fig:hgauge}
\end{figure}
These terms lead to  quadratically divergent Feynman
diagrams (Figure \ref{fig:hgauge})
that generate a mass term
\bea
{g^2 \over 16 \pi^2} \Lambda^2 h^\dagger h.
\eea
 Note that these diagrams are exactly the quadratically divergent
Standard Model gauge loops which we set out to cancel.
We have apparently gained nothing: we started with a
theory in which the Higgs was protected by a non-linearly
realized $SU(3)$ symmetry (under which $h$ shifts),
but then we added the term of Eq.~(\ref{eq:gaugestupid1}),
which completely and explicitly breaks the symmetry.
Of course, we necessarily have to break the shift symmetry
in order to generate gauge interactions
for $h$, but we must break the symmetry in a subtler
way to avoid quadratic divergences in the Higgs mass.
 
\underline{Second attempt:}
Let's couple $h$ to gauge fields through the more $SU(3)$-symmetric looking expression 
\bea
|g \left(\begin{array}{cc}
W_\mu &  \\  & 0 \end{array}\right) \phi\,|^2,
\label{eq:gaugestupid2}
\eea
where $W_\mu$ contains the three $SU(2)$ gauge bosons. (Really,
we write $|D_\mu \phi|^2$, where the covariant derivative involves only
$SU(2)$ gauge bosons. The two-gauge-boson-coupling is then
Eq.~(\ref{eq:gaugestupid2})).
This still generates a quadratically divergent contribution
to the Higgs mass. The diagram is the same as before except with external
$\phi$ fields, and gives
\bea
{g^2 \over 16 \pi^2} \Lambda^2 \ \phi^\dagger 
\left(\begin{array}{ccc} 1 & & \\ & 1 & \\ & & 0 \end{array}\right) \phi
={g^2 \over 16 \pi^2} \Lambda^2\, h^\dagger h + \cdots,
\eea
where the projection matrix
${\rm diag}(1,1,0)$ arises from
summing over the three $SU(2)$ gauge bosons running in the loop.
Not surprisingly, we got the same answer as before because we added
the same interactions, just using a fancier notation. 

\underline{Third attempt:}
Let us preserve $SU(3)$ by gauging the full $SU(3)$
symmetry, i.e. by adding $|D_\mu \phi|^2$, where now the covariant derivative
contains the 8 gauge bosons of $SU(3)$. Again we can write the same
quadratically divergent diagram and find
\bea
{g^2 \over 16 \pi^2} \Lambda^2\ \phi^\dagger 
\left(\begin{array}{ccc} 1 & & \\ & 1 & \\ & & 1 \end{array}\right) \phi
={g^2 \over 16 \pi^2} \Lambda^2 f^2,
\eea
which has no dependence on the Higgs field. The quadratic 
divergence contributes a constant term to the vacuum energy,
but no mass for the ``Higgs'' doublet $h$! Unfortunately, we have also lost  $h$:
the NGBs are ``eaten'' by the heavy $SU(3)$ gauge bosons
corresponding to the broken generators, i.e. they become the
longitudinal components of the gauge bosons.

We have now exhausted all possible ways of adding $SU(2)$ gauge
interactions to our simple toy model for $h$. The lesson
is that we can avoid the quadratically divergent
contribution to the Higgs mass by writing $SU(3)$-invariant gauge
interactions, the problem that remains is that our ``Higgs'' was
eaten. But this is easy to fix.

\begin{figure}[htb]
\vskip 0.2truein
\centerline{
{\includegraphics[width=2in]{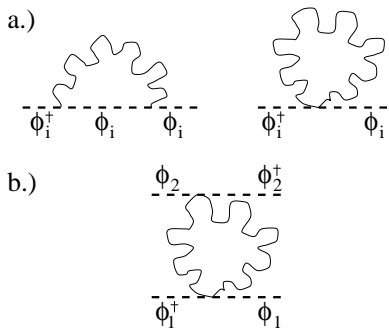}}}
\vskip- 0.1truein
\caption{a.) Quadratically divergent gauge loop contributions
which do not contribute to the Higgs potential, b.) log-divergent
contribution to the Higgs mass. }
\label{fig:gaugephi}
\end{figure}

\underline{Fourth attempt (successful):}
We use two copies of NGBs, $\phi_1$ and $\phi_2$, and add $SU(3)$
invariant covariant derivatives for both. We expect no quadratic
divergence for either of the NGBs, and only one linear combination
will be eaten. To see how this works in detail we parametrize
\bea
\phi_1 = e^{i \pi_1 /f}
\left(\begin{array}{c}   \\ f \end{array}\right)
\qquad
\phi_2 = e^{i \pi_2 /f}
\left(\begin{array}{c}   \\ f \end{array}\right),
\eea
where we have assumed aligned VEVs for $\phi_1$
and $\phi_2$, and  --  for simplicity -- identical symmetry breaking
scales $f_1=f_2=f$.
The Lagrangian is
\bea
\mathcal L = |D_\mu \phi_1|^2 + |D_\mu \phi_2|^2.
\eea
The two interaction terms produce two sets of  quadratically divergent
one-loop diagrams similar to those of the previous attempt 
(Figure \ref{fig:gaugephi}.a), which
give
\bea
{g^2 \over 16 \pi^2} \Lambda^2 \ (\phi_1^\dagger \phi_1
+\phi_2^\dagger \phi_2) = {g^2 \over 16 \pi^2} \Lambda^2 \ (f^2+f^2),
\eea
i.e. no potential for any of the NGBs. Moreover, only
one linear combination of $\pi_1$ and $\pi_2$
is eaten as there is only one set of hungry massive
$SU(3)$ gauge bosons.
A simple way to understand this result is to notice
that each set of diagrams involves only
one of the $\phi$ fields. Therefore the diagrams are the same as in the
theory with only one $\phi$, where all NGBs were eaten, and so neither  $\phi_1$ nor $\phi_2$
can get a potential. 

This reasoning doesn't apply once we consider diagrams 
involving both $\phi_1$ and $\phi_2$. For example, the diagram in
Figure  \ref{fig:gaugephi}.b gives
\bea
{g^4 \over 16 \pi^2} \log\left({\Lambda^2 \over \mu^2}\right) \ |\phi_1^\dagger \phi_2|^2,
\label{eq:logpotential}
\eea
which does depend on $h$ but is not quadratically divergent.
To calculate the Higgs dependence
we choose a convenient parametrization
\bea
\phi_1 &=&  exp\left\{i \left(\begin{array}{cc}  &k  \\ k^\dagger &  \end{array}\right) \right\}
exp\left\{i \left(\begin{array}{cc}  &h  \\ h^\dagger &  \end{array}\right) \right\}
\left(\begin{array}{c}   \\ f  \end{array}\right) \\
\phi_2 &=&  exp\left\{i \left(\begin{array}{cc}  &k  \\ k^\dagger &  \end{array}\right) \right\}
exp\left\{-i \left(\begin{array}{cc}  &h  \\ h^\dagger &  \end{array}\right) \right\}
\left(\begin{array}{c}   \\ f  \end{array}\right) .
\label{eq:parametrization}
\eea
The field $k$ can be removed by an $SU(3)$ gauge transformation, and
corresponds to the ``eaten" NGBs, while $h$ cannot simultaneously be removed from
$\phi_1$ and $\phi_2$, and is physical. In the following we will work in the unitary
gauge for $SU(3)$, where $k$ is rotated away.
Then we have 
\bea
\phi_1^\dagger \phi_2 &= &
\left(\begin{array}{cc}   0 & f \end{array}\right) 
exp\left\{-{2i\over f} \left(\begin{array}{cc}  &h  \\ h^\dagger &  \end{array}\right) \right\}
\left(\begin{array}{c}  0 \\ f  \end{array}\right) \nonumber \\
&=& \left[
f^2\left(\!\begin{array}{cc}  1&  \\  &1  \end{array}\!\right) 
- 2 f i  \left(\!\begin{array}{cc}  &h  \\ h^\dagger &  \end{array}\!\right) 
-2 \left(\!\begin{array}{cc}  h^\dagger h&  \\ & h^\dagger h   \end{array}\!\right) + \cdots
\right]_{33} \nonumber \\
&=& f^2 - 2 h^\dagger h + \cdots,
\eea
and we see that Eq.~(\ref{eq:logpotential}) contains a mass-squared for $h$ equal to
\begin{equation}
g^4/(16 \pi^2) \log\left({\Lambda^2 \over \mu^2}\right) f^2,
\end{equation}
which is  $\sim M_{weak}^2$
for $g$ equal to the $SU(2)$ gauge coupling and $f \sim$ TeV.

To summarize, the theory of two complex triplets which both break
$SU(3)\rightarrow SU(2)$ 
automatically contains a ``Higgs" doublet pseudo-NGB which
does not receive quadratically
divergent contributions to its mass. There are log-divergent
and finite contributions, and from these the
natural size for the ``Higgs" mass is  $f/4 \pi\sim M_{weak}$.
Thus our theory has three relevant scales which are separated
by loop factors, $\Lambda \sim 4 \pi f \sim (4 \pi)^2 M_{weak}$.

\subsection{Symmetry argument, collective breaking}

Let us understand the absence of one-loop quadratic divergence in the mass of $h$ 
using symmetries. The lesson we will learn is valuable
as it generalizes to other couplings, and so  provides a general recipe for
constructing little Higgs theories.

Without gauge interactions, our theory would consist of two
non-linear sigma fields, each representing the spontaneous breaking of
an $SU(3)$ global symmetry to $SU(2)$.  The coset  is thus $[SU(3)/SU(2)]^2$.
There are 10 spontaneously broken generators for this coset,  and
therefore 10 NGBs. The gauge couplings explicitly break
some of the global symmetries.
For example, the two gauge boson - two scalar coupling
\bea
\mathcal L \sim |g A_\mu \phi_1 |^2 + |g A_\mu \phi_2 |^2 
\label{eq:gaugecoup}
\eea
breaks the two previously independent $SU(3)$ symmetries
to the diagonal $SU(3)$ which is gauged. Thus only one of the
spontaneously broken symmetries is exact,
and  only one set of exact NGBs arises, the eaten ones.
The other linear combination, corresponding
to the explicitly broken axial $SU(3)$, gets a potential from loops.

However, as we saw before, there is no quadratically
divergent contribution to the potential.
This is easy to understand by considering the symmetries left
invariant by each of the terms in Eq.~(\ref{eq:gaugecoup}) separately.
Imagine setting the gauge coupling of $\phi_2$ to zero. 
Then the Lagrangian has 2 independent $SU(3)$ symmetries, one acting on $\phi_1$
(and $A_\mu$) and the other acting on $\phi_2$. Thus we now have two
spontaneously broken $SU(3)$ symmetries and therefore 
10 exact NGBs (5 of which are eaten). Similarly,
if the gauge coupling of $\phi_1$ is set to zero, there are again
two spontaneously broken $SU(3)$'s. Only in the presence of
gauge couplings for both $\phi_1$ and $\phi_2$ are the two $SU(3)$
symmetries explicitly broken to one $SU(3)$,  and only then can 
$h$ develop a potential. Therefore
any diagram which contributes to the $h$ mass must involve the
gauge couplings for both $\phi_1$ and $\phi_2$.
But there are no quadratically divergent one-loop diagrams involving
both couplings. 

This is the general mechanism employed by ``little Higgs" theories
\cite{ACG}:
\vskip.1in
\noindent {\it The ``little Higgs" is a
pseudo-Nambu-Goldstone boson of a spontaneously broken symmetry. This
symmetry is also explicitly broken but only ``collectively",
i.e. the symmetry is broken when two or more couplings in
the Lagrangian are non-vanishing. Setting any one of these couplings
to zero restores the symmetry and therefore the masslessness of the
``little Higgs".  }
\vskip.1in

We now know how to construct a theory with a naturally light scalar
doublet that couples to $SU(2)$ gauge bosons. To turn this into an extension
of the Standard Model we still need {\it i.} Yukawa couplings, {\it ii.}
hypercharge and color, and {\it iii.} a Higgs potential with a quartic coupling.

\subsection{Top Yukawa coupling}

The numerically most significant quadratic divergence stems from top quark
loops.  Thus the cancellation of the quadratic divergence associated with 
the top Yukawa is the most important. Let us construct a sector that guarantees this cancellation. The
crucial trick is to introduce $SU(3)$ symmetries into
the Yukawa couplings which are only broken collectively.
First, we enlarge the quark doublets into triplets $\Psi\equiv (t,b,T)$
transforming under the $SU(3)$ gauge symmetry. The quark
singlets remain the same, $t^c$ and $b^c$, except that we also
need to add a Dirac partner $T^c$ for $T$.
Note that we are using a notation in which all
quark fields are left-handed Weyl spinors, and 
the Standard Model Yukawa couplings are of the form $h^\dagger Q t^c$.
Let us change notation slightly to reflect the
fact that $t^c$ and $T^c$ mix,
and call them $t_1^c$ and $t_2^c$.
We can now write two terms that both look like they contribute
to the top Yukawa coupling%
\footnote{We do not write the couplings
$\phi_1^\dagger \Psi t_2^c$ and $\phi_2^\dagger \Psi t_2^c$ as they would
reintroduce quadratic divergences. They can be forbidden by global
$U(1)$ symmetries and are therefore not generated by loops.},
\bea
\mathcal{L}_{yuk} = \lambda_1 \phi_1^\dagger \Psi t_1^c 
+  \lambda_2 \phi_2^\dagger \Psi t_2^c.
\label{eq:topyuk}
\eea
To see what couplings for the Higgs arise we substitute the parametrization of
Eq.~(\ref{eq:parametrization}) and expand in powers of $h$. For simplicity,
let us also set $\lambda_1\equiv \lambda_2 \equiv \lambda/\sqrt{2}$. This will reduce
the number of terms we encounter because it preserves a parity
$1\leftrightarrow 2$, but the main points here are independent of this choice.
We find
\bea 
\mathcal L &\sim& {\lambda \over \sqrt{2}}
\left[ f T (t_2^c+t_1^c) + i h^\dagger Q (t_2^c-t_1^c) -\frac1{2f} h^\dagger h T (t_2^c+t_1^c) + \cdots 
\right] \nonumber \\
&=&\lambda f (1 -\frac{1}{2f^2} h^\dagger h) T T^c + \lambda h^\dagger Q t^c + \cdots, 
\label{eq:slhyuk}
\eea
where  the second line is written in terms of the linear combinations $T^c=(t_2^c+t_1^c)/\sqrt{2}$ and 
 $t^c=i(t_2^c-t_1^c)/\sqrt{2}$. 
 
 The last term of the second line in Eq.~(\ref{eq:slhyuk}) is the top Yukawa
coupling, and so we identify $\lambda=\lambda_t$. The Dirac fermion
 $T, T^c$ has a mass $\lambda_t f$ and a coupling to two
Higgs fields with coupling constant $\lambda_t/(2f)$.
The couplings and masses are related by the underlying $SU(3)$ symmetries.
\begin{figure}[htb]
\vskip 0.0truein
\centerline{
{\includegraphics[width=3in]{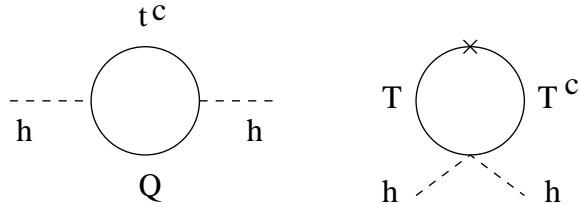}}}
\vskip- 0.0truein
\caption{The quadratically divergent contribution to the Higgs mass
  from the top loop is canceled by the $T$ loop.}
\label{fig:topcancel}
\end{figure}
 To see how the new fermion and its couplings to the Higgs
cancel the quadratic divergence from the top quark loop, we
compute the fermion loops including interactions to
 order $\lambda^2$. The two relevant diagrams
(Figure \ref{fig:topcancel}) give
 \bea
  { \lambda_t^2 \over 16 \pi^2}\ \Lambda^2 h^\dagger h + 
 { \lambda_t^2 f^2 \over 16 \pi^2} (1-\frac{h^\dagger h}{ f^2})\ \Lambda^2
+ \mathcal O(h^4) = 
{\rm const.} + \mathcal O(h^4).
 \eea
The quadratically divergent contribution to the Higgs
mass from the top and $T$ loops cancel!%
\footnote{
In order for the two cut-offs for the two loops to be identical, the new
physics at the cut-off must respect the $SU(3)$ symmetries. This is analogous to the
situation in SUSY where the boson-fermion cancellation
also relies on a supersymmetric regulator/cut-off.}

While this computation allowed us to see explicitly
that the quadratic divergence from
$t$ and $T$ cancel, the absence of a quadratic divergence to the Higgs mass
is much more naturally understood by analyzing the symmetries of the Lagrangian
for the $\phi_i$ fields, Eq.~(\ref{eq:topyuk}).
First note that the Yukawa coupling Lagrangian preserves one
$SU(3)$ symmetry, the gauge symmetry. The term proportional to $\lambda_1$ forces
symmetry transformations of $\phi_1$ and $\Psi$ to be aligned, and the term proportional to $\lambda_2$ 
also forces $\phi_2$ to transform like $\Psi$.
Thus, in the presence of both terms
the global symmetry breaking pattern is only $SU(3)/SU(2)$, with 5 NGBs which are
all eaten by the heavy $SU(3)$ gauge bosons.
However, if we set either of the $\lambda_i$ to zero,
the symmetry of Eq.~(\ref{eq:topyuk}) is enhanced to
$SU(3)^2$ because the $\phi_i$ can now
rotate independently. Thus, with either of the
$\lambda_i$ turned off, we expect two sets of NGBs. One
linear combination is eaten and the other is the ``little Higgs".

\begin{figure}[htb]
\vskip 0.0truein
\centerline{
{\includegraphics[width=1.5in]{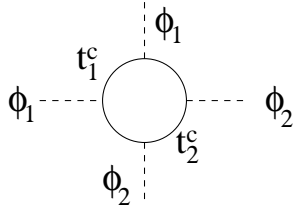}}}
\vskip- 0.0truein
\caption{A log divergent contribution to the Higgs mass
  from the top and $T$ loops proportional to $|\lambda_1 \lambda_2|^2$.}
\label{fig:topquartic}
\end{figure}

To understand radiative stability
of this result we observe that a contribution to the
Higgs potential can only come from
a diagram which involves both $\lambda_i$.
The lowest-order fermion diagram which involves
both $\lambda_i$ is the loop shown in Figure \ref{fig:topquartic}, which
is proportional to $|\lambda_1 \lambda_2|^2$.
You can easily convince yourself that you cannot draw a
diagram which contributes to the Higgs potential and is
proportional to only a single power of $\lambda_1 \lambda_2$.
This also follows from an argument using ``spurious" symmetries:
assign $t_1^c$ charge $1$ under
a $U(1)_1$ symmetry while all other fields are neutral.
The symmetry is broken by the
Yukawa coupling $\lambda_1$, but we can formally restore
it by assigning the ``spurion" $\lambda_1$
charge -1. Any effective operators which may be generated
by loops must be invariant under
this symmetry. In particular, operators which contribute
to the Higgs potential and do not
contain the fermion field $t_1^c$ can depend on the
spurion $\lambda_1$ only through $|\lambda_1|^2$.
Of course, the same argument shows that the dependence
on $\lambda_2$ is through $|\lambda_2|^2$
only. A contribution to the Higgs potential
requires both couplings $\lambda_1$ and $\lambda_2$
to appear and therefore the potential is proportional
to at least $|\lambda_1 \lambda_2|^2$, i.e.  it has at least four
coupling constants. But a one-loop diagram with 4 coupling
constants can at most be logarithmically
divergent, and therefore does not destabilize the Higgs mass.

In the explicit formulae above, we assumed for simplicity  that $f_1=f_2=f$ and
$\lambda_1=\lambda_2=\lambda_t/\sqrt{2}$. In the general case we find
\bea
m_T&=&\sqrt{\lambda_1^2 f_1^2 + \lambda_2^2 f_2^2} \\
\lambda_t &=& \lambda_1 \lambda_2 {\sqrt{f_1^2 + f_2^2} \over m_T}.
\eea
Note that the top Yukawa coupling goes to zero as either of the
$\lambda_i$ is taken to zero,
as anticipated from the $SU(3)$ symmetry arguments.
Furthermore note that the mass of the
heavy $T$ quark can be significantly lower than the
larger of the two $f_i$ if the corresponding
$\lambda_i$ is smaller than 1. This is a nice feature
because it allows us to take the heavy
gauge boson's masses large ($\gsim$ few TeV as required
by the precision electroweak constraints)
while keeping the $T$ mass near a TeV. Keeping the $T$
mass as low as possible is desirable
because the quadratic divergence of the top loop in the
Standard Model is cut off at the mass
of $T$.

\subsection{Other Yukawa couplings}

The other up-type Yukawa couplings may be added in exactly the same way.
We enlarge the $SU(2)$ quark doublets into triplets because of
the gauged $SU(3)$. Then we add two sets of Yukawa couplings which couple the
triplets to $\phi_1$ and $\phi_2$ and quark singlets $q^c_1$ and $q^c_2$.

In the Standard Model, Yukawa couplings for down-type quarks
arise from a different operator where the $SU(2)$ indices
of the Higgs doublet and the quark doublets are contracted using an
epsilon tensor (or, equivalently, the conjugate Higgs field
$h^c=i \sigma_2 h^\dagger$
is used). Before explicitly constructing this operator
from the quark and $\phi_i$ fields
note that even the bottom Yukawa coupling is too small
to give a significant contribution 
to the Higgs mass. The quadratically divergent one loop diagram in the
Standard Model yields
\bea
{\lambda_b^2 \over 16 \pi^2} \Lambda^2 \approx (30 {\rm\ GeV})^2
\eea
for $\Lambda \sim{\rm 10~TeV}$.
Therefore, we need not pay attention to symmetries and collective
breaking when constructing the down-type Yukawa couplings.
The Standard Model Yukawa is
\bea
\lambda_b \epsilon_{ij} h_i Q_j\, b^c.
\eea
To obtain the epsilon contraction from an $SU(3)$-invariant operator we write the Lagrangian term
\bea
{\lambda_b \over f}\,  \epsilon_{ijk} \phi_1^i \phi_2^j \Psi_Q^k\, b^c.
\eea
 Note that the $\epsilon_{ijk}$ contraction breaks both $SU(3)$ symmetries
(acting on the two scalar triplets $\phi_1$ and $\phi_2$) to the diagonal subgroup,
and therefore this operator does lead to a quadratic divergence.
But the quadratic divergence is harmless because of
 the smallness of the bottom Yukawa coupling.
 
 \subsection{Color and hypercharge}
 
 Color is added by simply adding $SU(3)_{\rm color}$
indices where we expect them from the Standard Model.
$SU(3)_{\rm color}$ commutes with
all the symmetry arguments given above,
therefore nothing significant changes.
 
Hypercharge is slightly more complicated.
The VEVs $\phi_i \propto (0,0,1)$ break the $SU(3)_{\rm weak}$
gauge group to $SU(2)$, i.e. no $U(1)$ hypercharge candidate is left.
Therefore, we gauge an additional $U(1)_X$. 
In order for the hypercharge of the Higgs to come out correctly,
we assign it $SU(3) \times U(1)_X$ quantum numbers
 \bea
 \phi_i=3_{-1/3}.
 \eea
 The combination of generators which is unbroken by $\phi_i \sim (0,0,1)$
 is 
 \bea
 Y={-1 \over \sqrt{3}} \, T^8 + X \quad {\rm where} \quad
 T^8 =  { 1 \over 2 \sqrt{3}}\ \left(\begin{array}{ccc}  1&& \\ &1&   \\  &&\!\!\!-2  \end{array}\right),
 \eea
 and $X$ is the generator corresponding to $U(1)_X$. This uniquely fixes the
 $U(1)_X$ charges of all quarks and leptons once their $SU(3)$ transformation 
 properties are chosen. 
 
 For example, the covariant derivative acting on $\phi_i$ is
 \bea
 D_\mu \phi = \partial_\mu \phi -
{1 \over 3}\, i g_X A^X_\mu \phi + i g A^{SU(3)}_\mu \phi.
 \eea
 Note that the $U(1)_X$ generator commutes with $SU(3)$, and the $U(1)_X$ gauge
 interactions do not change any of the symmetry arguments which we
 used to show that the Higgs mass is not  quadratically divergent at one loop.
 
 There are now three neutral gauge bosons, corresponding to the generators
 $T^3, T^8, X$. These gauge bosons mix, and
the mass eigenstates are the photon, the $Z$, and 
 a $Z'$.  The  $Z'$ leads to interesting modifications of
predictions for precision electroweak observables, as discussed in section \ref{sec:pheno}.

\subsection{Quartic Higgs coupling}

To generate a quartic Higgs coupling we want to write a potential $V(\phi_1,\phi_2)$
that {\it i.} contains no mass at order $f$ for the Higgs,
{\it ii.} contains a quartic coupling,
{\it iii.} preserves the ``collective" symmetry breaking of
the $SU(3)$'s: i.e. the quartic
coupling is generated by at least two couplings in $V$, and
if one sets either one of them to zero
the Higgs becomes an exact NGB. This last property is
what guarantees radiative stability,
that is, no $\Lambda^2$ contributions to the Higgs mass.

Writing down a potential which satisfies these properties
appears to be impossible for the pure $SU(3)$ model.
To see why it is not
straightforward, note that $\phi_1^\dagger \phi_2$
is the only non-trivial gauge invariant 
which can be formed from $\phi_1$ and $\phi_2$.
($\phi_1^\dagger \phi_1 =$ const $=\phi_2^\dagger \phi_2$
and $\epsilon^{ijk} \phi_i \phi_j \phi_k=0$).
But the $\phi_1^\dagger \phi_2$ invariant is a bad starting
point because it breaks the two $SU(3)$'s 
to the diagonal, and it is not surprising that generic functions
of $\phi_1^\dagger \phi_2$ always contain
a mass as well as a quartic. For example, we have
\bea
\phi_1^\dagger \phi_2 \sim f^2 - h^\dagger h + \frac1{f} (h^\dagger h)^2 + \cdots,
\eea
so that 
\bea
\frac1{f^{2n-4}} (\phi_1^\dagger \phi_2)^n \sim f^4 - f^2 h^\dagger h + (h^\dagger h)^2 + \cdots
\eea
By dialing the coefficient of this operator
we can either get a small enough
mass term or a large enough quartic coupling but not both.
Of course, we could try to tune two terms
with different powers $n$ such that the mass terms cancel
between them, but that tuning is
not radiatively stable.

There are two different solutions to the problem in the literature.
Both require enlarging the 
model and symmetry structure.
One solution, due to Kaplan and Schmaltz \cite{kaplanschmaltz},
involves enlarging the gauge symmetry to $SU(4)$ and
introducing four $\phi$ fields,  each transforming as a {\bf 4} of $SU(4)$.
The four $\phi$ fields break $SU(4) \rightarrow SU(2)$, yielding 
4 $SU(2)$ doublets. Two of them are eaten, and the other two are ``little
Higgs'' fields with a quartic potential similar to the quartic potential
in SUSY. 

The other solution, due to Skiba and Terning \cite{skibaterning},
keeps the $SU(3)$ gauge symmetry the same but
enlarges the global $SU(3)^2$ symmetry to $SU(3)^3$,  which is then
embedded in an $SU(9)$.
The larger symmetry also leads to two ``little''
Higgs doublets for which a quartic coupling can be written. 
Both of these solutions spoil some of the simplicity of the $SU(3)$
model, but they allow a large quartic coupling for the Higgs fields with
natural electroweak symmetry breaking. We refer the reader to the
original papers for details on these models.

A third option \cite{simplest} is simply to add a potential with a very small
coefficient. The resulting quartic coupling is then also very small, but as in the MSSM,
radiative corrections from the top loop  give a
contribution that can raise the Higgs mass above the experimental
bound of 114 GeV. 
\begin{figure}[htb]
\vskip 0.0truein
\centerline{
{\includegraphics[width=1.5in]{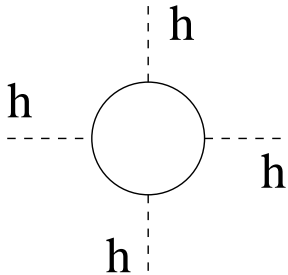}}}
\vskip- 0.0truein
\caption{The top loop contribution to the Higgs quartic coupling.}
\label{fig:toph4}
\end{figure}
Explicitly, below the $T$ mass the cancellations
in the top sector no longer occur and the diagram in Figure \ref{fig:toph4}
gives a quartic-term contribution
\bea
{3 \lambda_t^4 \over 16 \pi^2} \log({m_T^2 \over m_t^2}) \,
(h^\dagger h)^2,
\eea
which is too small by itself but does give successful electroweak symmetry
breaking when combined with a small tree-level contribution.
Since the tree-level term also contributes to  the Higgs mass-squared,
a moderate amount of tuning ($\sim$10\%) is required. While this is not
completely satisfactory, it is better than most other models of
electroweak symmetry breaking and certainly better than the MSSM
with gauge coupling unification, which requires tuning at the few \% level
or worse.

\subsection{The simplest little Higgs}

To emphasize the simplicity of the model, we summarize
the field content and Lagrangian of the ``simplest little Higgs''
\cite{simplest}, the $SU(3)$ model in which the
Higgs quartic coupling is predominantly generated from the top loop.

The model has an
$SU(3)_{\rm color} \times SU(3)_{\rm weak} \times U(1)_X$ gauge
group with three generations transforming as
\bea
\Psi_Q=(3,3)_{\frac13}\quad & \Psi_L=(1,3)_{-\frac13}\nonumber\\
d^c=(\bar 3,1)_{\frac13}\quad & e^c=(1,1)_1\nonumber  \\
2\times u^c=(\bar 3,1)_{-\frac23}\quad & n^c=(1,1)_0 
\eea 
The triplets $\Psi_Q$ and $\Psi_L$ contain the Standard Model
quark and lepton doublets, the singlets are $u^c, d^c, e^c, n^c$.%
\footnote{This fermion content is anomalous under the
extended electroweak gauge group. Anomalies may be canceled by
additional fermions which can be as heavy as $\Lambda$.
There are also charge assignments for which anomalies cancel
among the fields $\Psi_Q, \Psi_L, u^c, d^c, e^c, n^c$ alone
\cite{simplest,ottokong,ottokong2}.}
The $SU(3)_{weak} \times U(1)_X$ symmetry is broken by expectation
values for scalar fields $\phi_1=\phi_2=(1,3)_{-1/3}$.

The Lagrangian of the model contains the usual kinetic terms,
Yukawa couplings and a tree level Higgs potential
\bea
\mathcal L_{kin} & \sim& \Psi_{\!Q}^\dagger\, \slash\!\!\!\! D \Psi_Q + \cdots
+ |D_\mu \phi_1|^2 + \cdots \\
\mathcal L_{yuk} &\sim& \lambda^u_1 \phi_1^\dagger \Psi_Q u_1^c
+ \lambda^u_2 \phi_2^\dagger \Psi_Q u_2^c
+ {\lambda^d \over f} \phi_1 \phi_2 \Psi_Q\, d^c \nonumber \\
&&+ \lambda^n \phi_1^\dagger \Psi_L n^c
+ {\lambda^e \over f} \phi_1 \phi_2 \Psi_L\, e^c \\
\mathcal L_{pot} &\sim& \mu^2 \phi_1^\dagger \phi_2 \ .
\eea

Substituting the parametrization for the NGBs
\bea
\phi_1 = e^{i\Theta  {f_2\over f_1} }
\left( \begin{array}{l}
\!0\!\!  \\ \!0\!\! \\ \!f_1\!\! \end{array} \right), \
\phi_2= e^{-i \Theta {f_1\over f_2}}
\left( \begin{array}{l}
\!0\!\!  \\ \!0\!\! \\ \!f_2\!\!\end{array} \right) , \
\label{eq:phis}
\Theta = {\eta\over \sqrt{2}f} + \frac1f
\left( \begin{array}{cc} 
\!\!\begin{array}{ll} \!0\! & \!0\! \\ \!0\! & \!0\! \end{array} 
& \!\!h\! \\ h^\dagger  & \!\!0\! \end{array} \right)
\eea
where $f^2 = f_1^2+f_2^2$,
we can solve for the spectrum of heavy new gauge bosons
${W^+}', {W^0}', Z'$, fermions $T, U, C$ and
scalar $\eta$ \cite{simplest}.

\section{PRODUCT-GROUP MODELS}

\subsection{The Minimal Moose}
In this section, we describe two little Higgs theories whose structures are quite different than that of the $SU(3)$ model.  Taken together, the three models illustrate  different approaches to implementing the little Higgs mechanism economically.

First we describe the ``Minimal Moose'' model presented in Ref.~\cite{Arkani-Hamed:2002qx}.  The coset on which this model is constructed bears some similarity to that of the
chiral Lagrangian used to describe the low energy dynamics of QCD.  In the case of QCD, an approximate $SU(3)_L \times SU(3)_R$ chiral symmetry, which becomes exact in the limit of vanishing light quark masses, is spontaneously broken to its vector subgroup, so the coset is $(SU(3)_L\times SU(3)_R)/SU(3)_V$.  There is an octet of pseudo-NGB's associated with this spontaneous breaking, and these are understood to be the light mesons $(\pi^0, \pi^\pm, K^0, {\bar K}^0, K^{\pm}, \eta^0)$. 

The coset used for the Minimal Moose is $[SU(3)_L\times SU(3)_R/SU(3)_V]^4$, so their are {\it four} sets of NGB octets that appear in 
four sigma fields $\Sigma_{i}$, $i=1,2,3,4$.  The sigma fields transform under the global symmetries as
\begin{equation}
\Sigma_i \rightarrow L_i  \Sigma_i R_i^\dagger,
\end{equation}
where $L_i$ and $R_i$ are $SU(3)$ matrices.  The sigma fields can be parametrized in terms of the NGB's $\pi_i^a$, as
\begin{equation}
\Sigma_i=e^{2 i \pi_i^a T_a/f},
\end{equation}
where the $T_a$ are the generators of $SU(3)$, normalized so that ${\rm tr}(T_a T_b)={1 \over 2 }\delta_{ab}$.  

The idea is to identify some of these NGBs as  Higgs doublets responsible for electroweak symmetry breaking (this will turn out to be a two-Higgs-doublet model). 
As with the $SU(3)$ model, the challenge will be to give these fields gauge interactions, Yukawa couplings to fermions, and self interactions, in such a way that no quadratically divergent contributions to their masses squared arise at one loop.

\subsubsection{Gauge Interactions}
In this model, the electroweak group descends from a gauged  $SU(3)\times SU(2)\times U(1)$ subgroup of the original $SU(3)^8$ global symmetry. 
The sigma fields transform identically under this gauge group:  under  an $SU(3)$ gauge transformation $U$ and an $SU(2)\times U(1)$ transformation $V$, we have
\begin{equation}
\Sigma_i \rightarrow  U \Sigma_i V^\dagger.
\end{equation}
Here, $SU(2)\times U(1)$ gauge transformations are embedded in the $3 \times 3$ matrix $V$ in such a way that the $SU(2)$ transformation lives in the upper left $2 \times 2$ space of $V$, while the gauged $U(1)$ generator  is proportional to  ${\rm diag}(1,1,-2)$.   
We can take the Standard Model fermions to be charged under $SU(2) \times U(1)$ with their ordinary quantum numbers, and neutral under $SU(3)$.   Then to allow for couplings between the sigma fields and fermions, described below, the first two columns of $\Sigma$ are assigned $U(1)$ charge $-1/6$, while the last has charge $1/3$.

\begin{figure}
\centerline{
\includegraphics[width=3in]{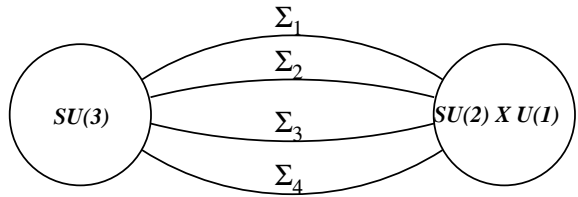}}
\caption{The Minimal Moose.}
\label{fig:moose}
\end{figure}

The gauge structure of the model can be depicted by the diagram of Fig.~(\ref{fig:moose}),
 in which there are two ``sites'' corresponding to the $SU(3)$ and $SU(2) \times U(1)$ gauge groups, and ``links" representing the sigma fields that transform under the gauge groups of both sites.  The Standard Model fermions live on the $SU(2) \times U(1)$ site.  A simple modification of this model is to have $SU(2) \times U(1)$ gauge groups at both sites, and we will return to this possibility in section \ref{subsec:othermodels}.

The sigma fields spontaneously break the $SU(3) \times SU(2) \times U(1)$ gauge symmetry to $SU(2)_{W} \times U(1)_{Y}$, where, as the subscripts suggest, the unbroken group is identified as the gauge symmetry of the standard electroweak theory.  
Similarly to what we did for the $SU(3)$ theory, we can assemble the NGBs in each sigma field into a matrix and make their decomposition under $SU(2)_W \times U(1)_Y$ manifest:
\begin{equation}
\pi_i=\pi_i^aT^a=\left( \begin{array}{cc}
 \phi_i+\eta_i /(2\sqrt{3})&h_i/\sqrt{2} \\
 h_i^\dagger /\sqrt{2} & -\eta_i/\sqrt{3} \\
 \end{array} \right), 
 \label{eq:pimatrix}
\end{equation}
where the real scalar $\eta_i$ is a singlet under $SU(2)_W \times U(1)_Y$, the real scalars $\phi_i=\phi_i^a \sigma^a/2$ transform as ${\bf 3_0}$, and the complex doublet $h_i$ transforms like the Standard Model Higgs, ${\bf 2_{1/2}}$, once we fix the normalization of $U(1)_Y$ charge appropriately.

By expanding the covariant derivatives in the  kinetic terms of the non-linear sigma model,
\begin{equation}
{\mathcal L}_{kin}={f^2 \over 4} \sum_{i=1}^4 {\rm tr} [(D_\mu \Sigma_i)^\dagger (D^\mu \Sigma_i)],
\end{equation}
one can calculate the masses of the octet of gauge bosons that become heavy when the gauge symmetry is spontaneously broken:  these  are $\sqrt{g_3^2+g_2^2} f$ for an electroweak triplet,  $g_3 f$ for a pair of doublets, and $\sqrt{g_3^2+g_1^2/3} f$ for a singlet.   
Here $g_1$, $g_2$, and $g_3$ are the $SU(3)$, $SU(2)$, and $U(1)$ gauge couplings, which determine the gauge couplings of the electroweak theory  according to 
\begin{eqnarray}
g' & = & {g_1 g_3 \over \sqrt{g_3^2+g_1^2/3}}\\
g & = & {g_2 g_3 \over \sqrt{g_2^2+g_3^2}}.
\end{eqnarray}  
Assuming that $f \sim$~TeV, we expect that the masses of these heavy gauge bosons will be somewhere near the TeV scale.

The gauge couplings  {\it explicitly} break the large $SU(3)^8$ global symmetry that we began with, and so the $\pi_i^a$ fields are exact NGBs only in the limit of vanishing gauge couplings.  
The only global symmetry that remains exact when the couplings are turned on is  $SU(3) \times SU(2) \times U(1)$ itself,  and the octet of NGBs associated with the spontaneous breaking of this symmetry to
$SU(2)_W \times U(1)_Y$ is eaten:  its members become the longitudinal components of the heavy gauge bosons.  The eaten fields appear in the  linear combination of $\pi_i$'s that shifts under $SU(3) \times SU(2) \times U(1)$ transformations associated
with the broken generators.  Because these transformations act identically on the various sigma fields, it is clear that  $\pi_1+\pi_2+\pi_3+\pi_4$ is the linear combination eaten (it shifts, while the linear combinations  orthogonal to it do not).

The uneaten NGBs acquire potentials from loops, but none receives quadratically divergent 
contributions to its mass at one loop.  In the limit that $g_3 \rightarrow 0$, an exact $SU(3)^4 \times SU(2) \times U(1)$ global symmetry is recovered.  This is spontaneously broken to $SU(2) \times U(1)$, ensuring a total of four exactly massless octets of NGBs (including the ones eaten).  If instead
$g_2$  and $g_1$ are turned off,  an $SU(3)^5$ global symmetry is recovered.  This is spontaneously broken to $SU(3)$, so all of the NGBs are massless once again.  Only when $g_3$ and at least one of $g_1$ and $g_2$ are turned on can these fields acquire mass, so a one-loop contribution to their masses must involve both $g_3$ and $g_2$, or both $g_3$ and $g_1$.  Contributions of this sort are at most logarithmically divergent.

The absence of one-loop quadratic divergences depends crucially on the enlarged gauge symmetry.  If only the diagonal $SU(2) \times U(1)$ is gauged, then 
there are no global symmetries to protect the masses of the scalars, and so they receive quadratically divergent contributions at one loop.  This is not surprising given that, in this case, there are no  extra particles around to cut off the divergences from ordinary gauge loops.  Once the extra $SU(3)$ is gauged, the heavy gauge bosons associated with the spontaneous breaking down to $SU(2)_W \times U(1)_Y$ appear in  one-loop diagrams whose quadratic divergences cancel those from diagrams with  massless gauge bosons running the loops.

\subsubsection{The Quartic Coupling}
A quartic coupling for the Higgs doublets can be generated by adding the following terms to the Lagrangian:
\begin{equation}
{\mathcal L}_\Sigma=\left( {f \over 2} \right)^4 {\rm tr}(c_1 \Sigma_1 \Sigma_2^\dagger \Sigma_3 \Sigma_4^\dagger) +\left( {f \over 2} \right)^4 {\rm tr}(c_2 \Sigma_1 \Sigma_4^\dagger \Sigma_3 \Sigma_2^\dagger) +{\rm h.c.},
\label{eq:plaquette}
\end{equation}
where for now we can imagine that $c_1$ and $c_2$ are numbers of order unity, which we take to be real for simplicity. 
Precisely where these operators come from is an interesting question, but 
there is no need to specify their ultraviolet origin to work out  their consequences in the effective theory.  In \cite{Arkani-Hamed:2002qx} it was shown that a particular arrangement of couplings of the sigma fields to fermions does generate these operators.

Let's consider the effects of ${\mathcal L}_\Sigma$, forgetting for the moment about the gauge couplings.  If only $c_1$, and not $c_2$, is turned on, the original  $SU(3)^8$ global symmetry is explicitly broken to $SU(3)^4$:   only those transformations satisfying $R_1=R_2$, $R_3=R_4$, $L_1=L_4$, and $L_2=L_3$ leave ${\mathcal L}_\Sigma$ invariant.  This $SU(3)^4$ global symmetry is spontaneously broken to $SU(3)$ by the sigma fields, giving three octets of of massless NGBs, one of which is eaten.  Since we started out with four octets, we expect  $c_1$ to give a potential to one of them.   

It is useful to define the linear combinations
\begin{eqnarray}
x & = &  {1\over \sqrt{2}} (\pi_1 -\pi_3) \\
y & = & {1\over \sqrt{2}} (\pi_2 -\pi_4) \\
z & =  & {1 \over 2} (\pi_1-\pi_2+ \pi_3-\pi_4),
\end{eqnarray}
and set the fourth linear combination, the eaten one, to zero.   The field $z$ is the linear combination that does not shift under any $SU(3)^4$ transformation preserved by $c_1$.  For instance, under infinitesimal symmetry transformations
 $R_1=R_2=1-2\sqrt{2}i\alpha/f$ and $L_1=L_4=1+2\sqrt{2} i \beta/f$, the changes in the fields are
\begin{eqnarray}
\delta x & =& \alpha +\beta+ \cdots  \nonumber \\
\delta y & =  & \alpha - \beta+\cdots  \nonumber \\
\delta z & = & {i \over 2f} \left( [\alpha +\beta,y]  -[\alpha - \beta, x]  \right)+\cdots,
\label{eq:mooseshift}
\end{eqnarray}
where terms of order $ x^2$, $y^2$, and $z$ have been neglected.  Both $x$ and $y$ shift, which tells us that no potential can be generated for them.  On the other hand, $z$ does not transform at zeroth order in the  fields, and so it can acquire a potential.  Expanding ${\mathcal L}_\Sigma$ for $c_2=0$, we find 
\begin{equation}
{\mathcal L}_{\Sigma}=- c_1 f^2 {\rm tr}\left( z-{i \over 2 f} [x,y] \right)^2+\cdots,
\label{eq:c1}
\end{equation}
where we have kept terms only up to order $z$, $x^4$, and $y^4$ in the interactions (it is easy to check that this expression is indeed invariant under the transformations of Eqn.~(\ref{eq:mooseshift})). As anticipated, only the $z$ octet acquires a mass-squared $\sim c_1 f^2$, while $x$ and $y$ remain massless.  In fact, after integrating out the $z$ octet, the potential for $x$ and $y$ vanishes completely: there is an exact cancellation between the ${\rm tr}[x,y]^2$ term already in Eqn.~(\ref{eq:c1}), and an identical term generated by the ${\rm tr}(z[x,y])$
coupling through exchange of the heavy $z$ octet.  This is all as it must be based on the symmetry considerations above.

When both $c_1$ and $c_2$ are turned on, we have
\begin{equation}
{\mathcal L}_{\Sigma}=- c_1 f^2 {\rm tr}\left( z-{i \over 2 f} [x,y] \right)^2- c_2 f^2 {\rm tr}\left( z+{i \over 2 f} [x,y] \right)^2.
\end{equation}
The crucial sign difference in the second term relative to the first  arises because the two terms in ${\mathcal L}_\Sigma$ are related by  $\Sigma_2 \leftrightarrow \Sigma_4$, which amounts to $\pi_2 \leftrightarrow \pi_4$, or $y \rightarrow -y$, with $x$ and $z$ left invariant.  Because of this sign difference, it is no longer true that the potential for $x$ and $y$ vanishes once $z$ is integrated out.  This makes sense: once $c_1$ and $c_2$ are both turned on, only a global $SU(3)^2$ symmetry is preserved, with each $L$ the same and each $R$ the same, and the only massless NGBs associated with the spontaneous breaking of this $SU(3)^2$ to $SU(3)$ are the eaten ones.  Thus, neither $x$ nor $y$ are exact  NGBs once $c_1$ and $c_2$ are both turned on, and a potential for them is possible at tree level.  

In particular, after integrating out $z$ we find the quartic term
\begin{equation}
{\mathcal L}_{quartic}={c_1 c_2 \over c_1+c_2} {\rm tr} [x,y] ^2.
\label{eq:quartic}
\end{equation} 
Because this term arises from an interplay between the $c_1$ and $c_2$ couplings, it vanishes if either is turned off.  The  tree-level potential produced  for $x$ and $y$ does not include mass terms, because neither $c_1$ nor $c_2$ can give mass to $x$ or $y$ by themselves,  and there is no tree-level diagram with intermediate $z$ that generates mass terms. Nor do the masses of the $x$ and $y$ fields receive quadratically divergent contributions at one loop, as there is no one-loop quadratically divergent diagram involving both $c_1$ and $c_2$. So the setup described here generates a tree-level quartic coupling for our pseudo-NGBs while keeping their masses-squared loop-suppressed relative to $f^2$.  This is exactly what we want:  if the quartic coupling suffered the same loop suppression as the mass terms, the hierarchy between $f$ and $v$ could not be realized without fine tuning.  

Note that  this mechanism of generating a tree-level quartic term for the NGBs  while protecting their masses would not work with fewer than four sigma fields.  For example, if we try using just three, and write down
\begin{equation}
{\mathcal L}_\Sigma=\left( {f \over 2} \right)^4 {\rm tr}(c_1 \Sigma_1 \Sigma_2^\dagger \Sigma_1 \Sigma_3^\dagger) +\left( {f \over 2} \right)^4{\rm tr}(c_2 \Sigma_1 \Sigma_3^\dagger \Sigma_1 \Sigma_2^\dagger) +{\rm h.c.}
\end{equation}
to generate a quartic coupling, 
we find that $c_1$ by itself breaks the global symmetry down to $SU(3)^2$. When this is spontaneously broken to $SU(3)$, it yields only the single of octet of massless NGBs that is eaten.

\subsubsection{The Top Yukawa Coupling} We demand that the top Yukawa coupling arises in such a way that no quadratically divergent contributions to the masses-squared of the Higgs doublets are induced, and the strategy will again be to ensure that each coupling preserves enough global symmetry to protect these masses.  The Standard Model fermions are not charged under $SU(3)$, so coupling the sigma fields to them requires us to work with $SU(3)$-singlet combinations, such as $\Sigma_1^T \Sigma_2^*$.  Now consider the Lagrangian terms
\begin{equation}
{\mathcal L}_t=\lambda f \left( \begin{array}{ccc}
0 & 0 & {u^c_3}' \\
 \end{array} \right) \Sigma_1^T \Sigma_2^* \chi
 + \lambda' f U^c U 
 +{\rm h.c.}
\label{eq:mmtop}
\end{equation}
Here, the third-generation electroweak-doublet  and singlet quarks $q_3=(u_3, d_3)$ and  ${u^c_3}'$ are joined by an extra vector-like electroweak-singlet quark $(U,U^c)$, and we have grouped some of these fermions into a triplet $\chi =(d_3, -u_3, U)$ (the ordering of $d_3$ and $u_3$ in $\chi$ is appropriate because it ensures that an $SU(2)_W$ transformation $V$, which sends $h \rightarrow V h$, acts on the upper two components of $\chi$ as $\chi \rightarrow V^* \chi$, as required for ${\mathcal L}_t$ to be gauge invariant).

As we will see shortly, these terms generate a Yukawa coupling for the top quark.  Moreover, they do so without generating quadratically divergent contributions to the $x$ and $y$ masses-squared at one loop, as can be seen by an argument similar to that used in discussing the top Yukawa coupling in the $SU(3)$ model: 
 in the absence of $\lambda'$, the term with coefficient ${\mathcal L}_t$ has an $SU(3)$ global symmetry under which $\chi$ transforms as a triplet, and this symmetry prevents $x$ and $y$ from acquiring mass.
So both couplings are required to generate masses for $x$ and $y$, and the rephasing symmetries of ${\mathcal L}_t$ indicate that these couplings must appear as $|\lambda|^2$ and $|\lambda'|^2$ in any radiatively generated operators involving only $\Sigma$.  Thus any one-loop diagram that contributes to
the $x$ and $y$ masses must involve four coupling constants, and can be at most logarithmically divergent.

The absence of quadratic divergences can also be verify explicitly, using the Coleman-Weinberg potential.  At one loop, the effective potential generated for $x$ and $y$ by fermion loops can be calculated by turning on background values for the sigma fields, and calculating the fermion mass matrix $M(\Sigma)$ in this background.  The quadratically divergent piece of the effective potential is then proportional to ${\rm tr} M(\Sigma)^\dagger M(\Sigma)$.   In our case, we have
\begin{equation}
{\mathcal L}_t= \left( \begin{array}{cccc}
0 & 0 & {u^c_3}' & U^c
 \end{array} \right) 
 M(\Sigma) \chi 
 +{\rm h.c.}, 
\end{equation}  
where the mass matrix $M(\Sigma)$ has the form
\begin{equation} 
M(\Sigma)=P \left( \begin{array}{ccc}
\multicolumn{3}{c} { \lambda f \Sigma_1^T \Sigma_2^*} \\
0 & 0& \lambda' f
\end{array} \right),
\label{eq: fermionmatrix}
\end{equation}
with $P={\rm diag}(0,0,1,1)$.   The unitarity of the $\Sigma$'s and the cyclic property of the trace then leads to
\begin{equation}
{\rm tr} M(\Sigma)^{\dagger} M(\Sigma)  = f^2 (|\lambda|^2+ |\lambda'|^2).
\end{equation}
There is no dependence on $x$ or $y$, and so as expected, no quadratically divergent contributions to their masses at one loop.  This technique also allows one to check explicitly that the gauge interactions and ${\mathcal L}_\Sigma $ do not generate quadratically divergent contributions at one loop either; see \cite{Arkani-Hamed:2002qx} for details.

To explore the coupling of the light scalars to fermions, we expand $x$ and $y$ in terms of $h_{x,y}$, $\eta_{x,y}$, and $\phi_{x,y}$ as in Eqn.~(\ref{eq:pimatrix}), and find that ${\mathcal L}_t$ includes the terms
\begin{equation}
\lambda' f U^c U+ \lambda f {u_3^c}' U +i {\lambda}
u_3^c q_3 (h_x-h_y) +{\rm h.c.}
\end{equation}
One linear combination of ${u_3^c}'$ and $U^c$ marries $U$ to make a Dirac fermion with mass $f \sqrt{|\lambda|^2+ |\lambda'|^2}$, while the orthogonal combination, $u_3^c$, is massless in the absence of electroweak symmetry breaking.
Integrating out the heavy fermion, we are left with the Yukawa coupling
\begin{equation}
\lambda_t u_3^c q_3 (h_x-h_y)/\sqrt{2},
\end{equation}
where the coefficient is
\begin{equation}
\lambda_t={\sqrt{2} \lambda \lambda' \over \sqrt{|\lambda|^2+ |\lambda'|^2}}.
\end{equation}
Thus an unsuppressed top Yukawa coupling is generated, without quadratically divergent contributions to the Higgs mass at one loop. 
The Yukawa couplings for the remaining Standard Model fermions are small enough that their quadratically divergent contributions to the Higgs mass-squared are 
numerically unimportant for the relatively low cut-off $\Lambda \sim 10~{\rm TeV}$ we have in  mind.    This means that to give masses to these fermions, we are free to write down couplings
similar to the first term in Eqn.~(\ref{eq:mmtop}), except without introducing extra vector-like fermions analogous to $(U,U^c)$.

 Note that ${\mathcal L}_t$ breaks the original $SU(3)^8$ global symmetry to $SU(3)^5 \times [SU(2) \times U(1)]^2$, which is spontaneously broken to $SU(3)^2 \times SU(2) \times U(1)$, giving
three full octets of massless NGBs plus one  $(\eta, \phi)$ set.  One full octet is eaten, and another becomes massive at tree level due to ${\mathcal L}_\Sigma$, leaving a full octet plus an additional  $(\eta, \phi)$ set as light fields that acquire {\it no} potential due to ${\mathcal L}_t$.
This tells us that of the fields  $x$ and $y$, top loops contribute to the mass of only one combination of Higgs doublets ($h_x -h_y$), and do not contribute to any $\eta$ or $\phi$'s masses.

\subsubsection{Electroweak Symmetry Breaking}  All of the light scalars in $x$ and $y$ receive positive contributions to their masses-squared once one-loop contributions induced by ${\mathcal L}_\Sigma$ and gauge interactions are taken into account.  As just discussed, the triplets and singlets do not receive contributions to their potentials due to ${\mathcal L}_t$, but the linear combination $h_x -h_ y$  {\it does} receive a one-loop contribution to its mass from the coupling to fermions, and this contribution is negative.  It is possible that this negative contribution is large enough to force a vev for $h_x -h_y$, causing electroweak symmetry breaking.

To address the important question of the stability of electroweak symmetry breaking, let us expand the quartic coupling of Eqn.~(\ref{eq:quartic}) in terms of the Higgs doublets.  We find
 \begin{equation}
 {\mathcal L}_{quartic}={c_1 c_2 \over 4 (c_1+c_2)} \left(  (h_x^\dagger h_y-h_y^\dagger h_x)^2 +{\rm tr}(h_y h_x^\dagger-h_x h_y^\dagger)^2 \right),
 \end{equation}
 where we've set $\eta_{x,y}$ and $\phi_{x,y}$ to zero.  Defining new doublets $h_1=(h_y-i h_x)/\sqrt{2}$ and $h_2=(h_x-i h_y)/\sqrt{2}$, these terms become
 \begin{equation}
{\mathcal L}_{quartic} =-{c_1 c_2 \over 4 (c_1+c_2)} \left(  (h_1^\dagger h_1-h_2^\dagger h_2)^2 +{\rm tr}(h_1 h_1^\dagger-h_2 h_2^\dagger)^2 \right),
\end{equation}
which, when restricted to neutral components, has the same form as the quartic potential of the supersymmetric Standard Model.  This quartic potential
has a flat direction along $h_1=h_2$, which must be stabilized by the quadratic terms in the potential.  These are of the form $m_1^2 (|h_x|^2+|h_y|^2)+m_2^2 |h_x-h_y|^2$, where $m_2^2 <0$ comes from top loops.   Written in terms of $h_1$ and $h_2$,
the quadratic terms become
\begin{equation}
(m_1^2+ m_2^2)(|h_1|^2+|h_2|^2)+m_2^2 (h_1^\dagger h_2+h_2^\dagger h_1).  
\end{equation}
Now, for electroweak symmetry to occur at all, we need the origin to be unstable, which requires 
\begin{equation}
(m_1^2+ m_2^2)^2-(m_2^2)^2 < 0.
\end{equation}
Because $m_2^2$ is negative, this condition is easily satisfied.   But if it is satisfied, then  the quadratic terms in the potential do not stabilize the $h_1=h_2$ flat direction, and the potential is unbounded from below.  

This means that the sigma field couplings of Eqn.~(\ref{eq:plaquette}) need to be modified.  The modification suggested in \cite{Arkani-Hamed:2002qx} is to give a non-trivial matrix structure
to the couplings $c_1$ and $c_2$:  $c_1 \rightarrow c_1 {\bf I}+ \epsilon_1 {\bf T_8}$ and  $c_2 \rightarrow c_2 {\bf I}+ \epsilon_2 {\bf T_8}$.  The coupling $\epsilon_1$ breaks
the global symmetry of the term with coefficient $c_1$ from $SU(3)^4$ to $SU(2) \times U(1) \times SU(3)^3$.  It is true that, since this global symmetry is spontaneously broken to $SU(2) \times U(1)$, the $\epsilon_1$ term by itself would still leave three massless octets of NGBs.  The point, though, is that the three octets protected when $c_1$ alone is turned on are not exactly the same as the three octets protected when $\epsilon_1$ alone is turned on, and when both are turned on, some of the states contained in $x$ and $y$ acquire mass.  By explicit calculation, one finds the Lagrangian term
\begin{equation}
{\sqrt{3} \over 8} f^2 \Im (\epsilon_1 - \epsilon_2) i (h_y^\dagger h_x -h_x^\dagger h_y),
\end{equation}
which is proportional to $|h_1|^2-|h_2|^2$.  Now that the masses-squared for $h_1$ and $h_2$ are split, it is possible to have stable electroweak symmetry breaking. For this term to have the right size, $ \Im(\epsilon_1 - \epsilon_2)$ should be small, $\sim 10^{-2}$.  However, since  $\epsilon_1$ and $\epsilon_2$ encode the effects of additional sources of symmetry breaking, it is natural for them to be suppressed.  
  
To summarize, the Minimal Moose has  two Higgs doublets in its scalar sector, along with a complex triplet  and a complex singlet, all with masses near the weak scale.  Around the TeV scale,
there is an additional octet of scalars, along with an electroweak-singlet vector-like quark, and an octet of heavy gauge bosons.  This particle content is sufficient for
cutting off the one-loop quadratic divergences associated with the Higgs doublets' gauge, Yukawa, and self couplings at around the TeV scale.
  
 \subsection{The Littlest Higgs}
 The third and final model that we will discuss in detail is the ``Littlest Higgs'' model of Ref.~\cite{Arkani-Hamed:2002qy}, which is constructed using an $SU(5)/SO(5)$ coset.  We
 can imagine that the $SU(5)$ global symmetry is broken by  the vacuum expectation value of a scalar $\Sigma$ transforming as a two-index symmetric tensor, or ${\bf 15}$ of $SU(5)$.  Let's take this vev
 to have the form
 \begin{equation}
 \langle  \Sigma \rangle =\left( \begin{array}{ccc}
\quad & \quad  & \openone \\
 \quad & 1 &  \quad \\
 \openone & \quad  & \quad 
 \end{array} \right),
 \end{equation}
where $\openone$ is the $2 \times 2$ identity matrix, and entries left blank vanish.
Under an SU(5) transformation $U=e^{i \theta_a T_a}$, we have $\Sigma \rightarrow U \Sigma U^T$, which means that the
ten unbroken generators satisfy
 \begin{eqnarray}
 {\overline T}_a  \langle  \Sigma \rangle+\langle  \Sigma \rangle {\overline T}_a^T & = &  0. \quad \quad (\text{unbroken})
 \label{eq:unbroken}
 \end{eqnarray}
 These are the 10 generators of $SO(5)$.  
The 14 remaining broken generators  satisfy
  \begin{eqnarray}
{\hat T}_a  \langle  \Sigma \rangle-\langle  \Sigma \rangle {\hat T}_a^T & = & 0, \quad \quad (\text{broken})
\label{eq:broken}
 \end{eqnarray}
and in constructing the non-linear sigma model, we keep only the fluctuations of $\Sigma$ around its vev in these broken directions:
\begin{equation}
\Sigma =e^{i\pi_a {\hat T}_a /f}  \langle \Sigma  \rangle e^{i \pi_a {\hat T}_a^T/f} =e^{2 i \pi_a {\hat T}_a/f}   \langle \Sigma  \rangle, 
\end{equation}
where the last equality follows from Eqn.~(\ref{eq:broken}).  

It is straightforward to show that, in light of Eqn.~(\ref{eq:broken}),
the matrix of NGBs may be written as
\begin{equation}
\pi= \pi_a {\hat T}_a =\left( \begin{array}{ccc}
 \chi+\eta/( 2\sqrt{5}) & h^*/\sqrt{2} & \phi^\dagger \\
h^T /\sqrt{2} & -2 \eta/\sqrt{5} & h^\dagger /\sqrt{2} \\
 \phi & h  / \sqrt{2} &  \chi^T+\eta/( 2\sqrt{5})
 \end{array} \right),
 \label{eq:littlestmatrix}
\end{equation}
where $\chi=\chi^a \sigma^a/2$ is a Hermitian, traceless $2 \times 2$ matrix, $\eta$ is a real singlet, $h$ is a complex doublet, and $\phi$ is
a $2 \times 2$ symmetric matrix.  The various coefficients are chosen so that the kinetic term of the non-linear sigma model,
\begin{equation}
{\mathcal L}_{kin}={f^2 \over 8}{\rm tr}[(D_\mu \Sigma)^\dagger(D^\mu \Sigma)].
\end{equation}
yields canonically normalized kinetic terms for the NGBs.

\subsubsection{Gauge Interactions}
As was done in the previous models, we now gauge a subgroup of the global symmetry, and as before, we do this in such a way that each gauge coupling by
itself preserves enough of the global symmetry to ensure that the Higgs doublet (a single doublet in this model) remains an exact NGB.  The gauge group is taken
to be $[SU(2) \times U(1)]^2$, embedded in $SU(5)$ in such that the the gauged generators for the  two $SU(2)$'s  are
\begin{equation}
Q_1^a =\left( \begin{array}{cc}
 -{\sigma^a}^*/2 & \quad   \\
 \quad & \quad
 \end{array} \right) \hspace{.5in} {\rm and} \hspace{.5in} Q_2^a =\left( \begin{array}{cc}
 \quad& \quad \\
  \quad &\quad  \sigma^a/2
 \end{array} \right),
\end{equation}
so the first $SU(2)$ acts on the first two indices, and the second acts on the last two.  The gauged generators for the two $U(1)$'s are 
 $Y_1={\rm diag}(-3,- 3, 2, 2, 2)/10$ and   $Y_2={\rm diag}(-2, -2, -2, 3, 3)/10$, respectively.  
Because the trace of the product of these generators is non-vanishing,
kinetic mixing between the two $U(1)$ gauge bosons will arise at loop level.
Re-diagonalizing the gauge kinetic terms modifies the couplings of
the $W$ and $Z$ to fermions. But aside from unobservable re-scalings
of gauge couplings, physical effects of the kinetic mixing
vanish in the limit of unbroken standard model gauge symmetry.
Therefore contributions to precision electroweak observables are
suppressed by $1/(16 \pi^2) (M_Z/M_{Z'})^2 \sim 10^{-4}$, evading
all constraints.  

The linear combination $Q_1^a+Q_2^a$  satisfies Eqn.~(\ref{eq:unbroken}), and generates the unbroken symmetry that we identify as $SU(2)_W$ of the Standard Model.
Similarly, the linear combination $Y_1+Y_2$ also satisfies Eqn.~(\ref{eq:unbroken}), and generates the unbroken symmetry that we identify as $U(1)_Y$.  The orthogonal combinations
are broken, and by expanding ${\mathcal L}_{kin}$ one can check that the heavy $SU(2)$ and $U(1)$ gauge bosons have masses $M_{W'}  =  f \sqrt{g_1^2+ g_2^2}/2$ and $M_{B'} =   f \sqrt{({g_1'}^2+{g_2'}^2)/20}$, respectively.
Here $g_1$ and $g_2$ are the gauge couplings of the two $SU(2)$'s, and $g_1'$ and $g_2'$ are the gauge couplings of the two $U(1)$'s.   These determine the Standard Model gauge couplings
\begin{eqnarray}
g & = & {{g_1 g_2 \over \sqrt{g_1^2+g_2^2}}}\\
g' & = & {{g_1'  g_2' \over \sqrt{ {g_1'}^2+{g_2'}^2}}},
\end{eqnarray}
where the  $U(1)$ charge is normalized so that $Y=1/2$ for the doublet $h$ ({\it i.e.} $Y=Y_1+Y_2$).  Note that the mass of the $B'$ is somewhat suppressed.  As discussed in section \ref{sec:pheno},  the effects of this particle modify $M_W/M_Z$ and the couplings of the $Z$ to fermions,  so the relative smallness  of $m_{B'}/f$ leads to important constraints on the model.  

When $[SU(2) \times U(1)]^2$ is broken to its diagonal subgroup, the $\eta$ and $\chi$ fields of the $\pi$ matrix in Eqn.~(\ref{eq:littlestmatrix}) are eaten, leaving only $h$ and $\phi$, which transform as ${\bf 2_{1/2}}$ and ${\bf 3_1}$ under $SU(2)_W \times U(1)_Y$.  What can we say about the potential for these fields generated by gauge loops?  Let's first imagine that all of the gauge couplings vanish, except for $g_2$.  In this case, the original $SU(5)$ global symmetry is explicitly broken to $SU(3) \times SU(2) \times U(1)$, where the $SU(3)$ acts on the first three indices, and the $SU(2)$ acts on the last two.   This is spontaneously broken to the $SU(2) \times U(1)$ of the electroweak group, and so in this limit, there are eight NGBs:  these include the four that are eaten when the full $[SU(2) \times U(1)]^2$ is gauged, and the four that make up the Higgs doublet $h$.  The Higgs doublet $h$ shifts under part of the $ SU(3)$, and so its forbidden from picking up a potential.  The six real scalars in $\phi$, on the other hand, are not protected by the global symmetry and are allowed to pick up a potential.  The same symmetry argument applies when only $g_2'$ is turned on.  Finally, when $g_1$ or $g_1'$ are turned on, a different
$SU(3) \times SU(2) \times U(1)$ is preserved (this time with the $SU(3)$ acting on the last three indices), and this symmetry is also enough to keep the Higgs massless.    The fact that more than one coupling is required to break enough of the global symmetry to let $h$ acquire a potential tells us that even when all couplings are turned on, the Higgs does not receive quadratically divergent contributions to its mass-squared at one loop.  

On the other hand, $\phi$ does pick up a quadratically divergent mass squared at one loop, and after integrating $\phi$ out, a tree-level quartic 
potential is generated for $h$.  This is very similar to what happened in
the Minimal Moose model with the introduction of ${\mathcal L}_{\Sigma}$, except that in the present model, no extra terms for $\Sigma$ are required:  the gauge interactions by themselves generate a quartic coupling for the Higgs.   This can be verified by studying the quadratically divergent piece of the Coleman-Weinberg potential, $V=\Lambda^2 {\rm tr} M^2(\Sigma)/(16 \pi^2) \sim f^2  {\rm tr} M^2(\Sigma)$, where ${\rm tr} M^2(\Sigma)$ is  the trace of the gauge boson mass-squared matrix in the presence of a background value for $\Sigma$.  It is calculated from ${\mathcal L}_{kin}$ as
\begin{equation}
{\rm tr} M^2(\Sigma)={f^2 \over 2}\sum_{i=1,2}
\left( {g_i'}^2 {\rm tr}[(\Sigma^\dagger Y_i)^* (\Sigma Y_i)]+{g_i}^2{\sum_a}  {\rm tr}[(Q_i^a \Sigma)^* (Q_i^a \Sigma) ] \right),
\end{equation}
where terms that do not depend on $\Sigma$ have been dropped.      
Expanding this expression gives
\begin{equation}
{\rm tr} M^2(\Sigma)  =  (g_1^2+{g_1'}^2) {\rm tr}({K_-}^\dagger K_-) + (g_2^2+{g_2'}^2) {\rm tr} ({K_+}^\dagger K_+) +\cdots, 
\label{eq:su5cwgauge}
\end{equation}
where $K_\pm=\phi \pm {i \over 2 f} h h^T$.
As claimed, $\phi$ picks up a quadratically divergent mass squared at one loop, but $h$ does not.  In fact, the potential for $h$ vanishes entirely once $\phi$ is integrated out if only the first term is present, or only the second term.  In the presence of both terms, however, integrating out $\phi$ generates the quartic term
\begin{equation}
{\mathcal L}_{quartic}=-c {(g_1^2+{g_1'}^2)(g_2^2+{g_2'}^2)\over g_1^2+{g_1'}^2+ g_2^2+{g_2'}^2}|h h^\dagger|^2,
\label{eq:su5quartic}
\end{equation}
where $c$ is a coefficient of order unity that encodes the details of how the quadratic divergences are cut off in the full UV-completed theory.  

\subsubsection{The Top Yukawa Coupling} In this model, there are a number of approaches to generating a top Yukawa coupling while protecting the Higgs mass.  Here we describe a setup that introduces the minimal number of additional fermions.    To avoid generating quadratically divergent contributions to the Higgs mass-squared at one loop, we require that the coupling of the top quark to $\Sigma$ respect one of the $SU(3)$ global symmetries under which $h$ shifts.  Suppose that we take the third generation quark doublet $q_3 =(u_3, d_3)$ to transform under $SU(2)_1$.  Then to achieve an $SU(3)$-symmetric coupling, we proceed as we did for the Minimal Moose: we add an extra electroweak-singlet vector-like fermion  $({U,U^c})$, form a triplet $\chi=(d_3, - u_3, U)$, and couple this triplet to $\Sigma$ and the electroweak singlet ${u_3^c}'$ an an $SU(3)$-symmetric fashion. 

Consider the Lagrangian terms 
\begin{equation}
{\mathcal L}_t=\lambda f \epsilon_{ijk} \chi_i \Sigma_{j4} \Sigma_{k5} {u_3^c}' + \lambda' f UU^c +{\rm h.c.} 
\label{eq:littlestfermion}
\end{equation}
The indices $i,j,k$ run over the values $1,2,3$, so the first term is an $SU(3)$-invariant antisymmetric contraction of three triplets.  It is also invariant under the gauged $SU(2)_2$, as required: for each $i$ and $j$, the combination $\Sigma_{i4} \Sigma_{j5}-\Sigma_{i5} \Sigma_{j4}$ appears after the sum is carried out, and this is an $SU(2)_2$-invariant antisymmetric contraction of two doublets.  The $U(1)_1 \times U(2)_2$ charges of the fermions are chosen so that the above terms are neutral.  For example, if we take $(1/6, 0)$ to be the charges of $q_3$, then we have ${u_3^c}'(-7/15, -1/5)$ and $U(2/3,0)$.  Anomalies associated with the broken generators of the extended gauge group are assumed to be canceled by heavy fermions.

In the presence of $\lambda$ alone, a linear combination of $u_3$ and $U$ remains massless even after electroweak symmetry breaking, but the second term makes this mode heavy.
When we expand ${\mathcal L}_t$ we find
\begin{equation}
{\mathcal L}_t = i \sqrt{2} \lambda  q_3 h {u_3^c}' +f U(\lambda {u_3^c}' + \lambda' U^c) +{\rm h.c.},
\end{equation}
where we have only kept terms up to linear order in $h$.  Inspection of the second term shows that $U$ marries a linear combination of ${u_3^c}'$ and $U^c$ to become a Dirac particle with mass $f \sqrt{|\lambda|^2+|\lambda'|^2}$.
Once this particle is integrated out, the orthogonal linear combination, $u_3^c$, appears in the Yukawa coupling
\begin{equation}
{\mathcal L}_t ={\sqrt{2} \lambda \lambda' \over \sqrt{|\lambda|^2+|\lambda'|^2}} q_3 h u_3^c +{\rm h.c.}
\end{equation} 
As discussed earlier, the Yukawa couplings for the remaining Standard Model fermions are small enough that we needn't worry about the quadratic divergences they generate.
We can produce these couplings by writing down terms 
similar to the first term in Eqn.~(\ref{eq:littlestfermion}), except without introducing extra vector-like fermions analogous to $(U,U^c)$.

We can  verify that the couplings in ${\mathcal L}_t$ do not generate quadratically divergent contributions to the Higgs mass-squared at one loop by calculating the quadratically divergent piece of the Coleman-Weinberg potential.  Defining $\phi_i=\epsilon_{ijk} \Sigma_{j4} \Sigma_{k5}$, the fermion mass matrix in a $\Sigma$ background is 
\begin{equation}
 M(\Sigma)=  f\left( \begin{array}{cc}
\lambda  \phi_1 & 0\\
\lambda  \phi_2 & 0\\
\lambda  \phi_3 & \lambda' 
 \end{array} \right).
\end{equation}
The quadratically divergent piece of the Coleman-Weinberg potential is then proportional to 
\begin{equation}
{\rm tr} M(\Sigma)^\dagger M(\Sigma)  =  f^2(|\lambda|^2 \sum_i |\phi_i|^2+|\lambda'|^2) 
 =  -|\lambda|^2 {\rm tr}({K_-}^\dagger K_-) , \label{eq:su5cwfermion}
\end{equation}
where for the last equality we have kept only terms up to order $\phi^2$ and $h^4$.  This term has the same form
as the second term in Eqn.~(\ref{eq:su5cwgauge}), which should come as no surprise given that $\lambda$ and $g_2$ respect the same global
symmetry.  When we integrate out $\phi$, the contributions to the potential arising from Eqn.~(\ref{eq:su5cwfermion}) will modify the quartic coupling
obtained in Eqn.~(\ref{eq:su5quartic}):  in the end, the quartic coupling depends on unknown order-one coefficients that are dependent on UV physics,
along with the gauge couplings and $\lambda$.    

Although ${\mathcal L}_t$ does not generate a quadratically divergent contribution to the Higgs mass-squared at one loop, it does generate a logarithmically divergent contribution at this level.  Moreover, this contribution is negative, and if its magnitude is sufficient large, it can overcome the positive contributions from gauge and self-interactions and cause electroweak symmetry breaking.

We have seen that the $SU(5)/SO(5)$ coset allows for a very economical implementation of the Little Higgs mechanism.  In this model, the Higgs doublet of the Standard Model is the only anomalously light pseudo-NGB.  The other scalars, which form an electroweak triplet, have a mass $\sim f \sim~{\rm TeV}$.  The other new states with masses around this scale are one electroweak-singlet vector-like quark (for the model of the top Yukawa coupling presented here), and weak-triplet and singlet heavy gauge bosons.    An especially interesting feature of this model is that the gauge interactions of the non-linear sigma field are by themselves sufficient for generating a quartic coupling for the Higgs doublet.  

\subsection{Other Models}
\label{subsec:othermodels}
In the next section we will discuss various effects that give rise to  precision electroweak constraints on little Higgs theories.  An order-$v^2/f^2$ modification of $M_W/M_Z$ arises from tree-level exchange of the heavy electroweak-singlet gauge boson.  Depending on the theory, corrections to $M_W/M_Z$ may also come from a triplet VEV, or from dimension-six operators in the expansion of  the non-linear sigma model kinetic term. Finally, integrating out the heavy gauge bosons also modifies the couplings of the light Standard Model gauge bosons to fermions and generates four-fermion operators in the effective theory.  Since these effects impose serious constraints on the models described so far, we now briefly describe alternative models in which the constraints are less severe.  

\subsubsection{Losing the Triplet}
A variation of the Littlest Higgs model based on the coset $SU(6)/Sp(6)$
was constructed in Ref.~\cite{Low:2002ws}.  This  breaking pattern gives rise to fourteen NGBs:  four are eaten due to the gauge symmetry breaking $[SU(2)\times U(1)]^2 \rightarrow SU(2) \times U(1)$, 8 appear in two Higgs doublets, and 2 appear in a neutral complex singlet.  The singlet's mass-squared receives quadratically divergent contributions at one loop, while the doublets remain light, and integrating out the singlet generates a quartic coupling for the Higgs doublets.  Having a extra singlet  instead of an extra triplet  allows for smaller values of $f$, as discussed in the following section.

\subsubsection{Models with Custodial Symmetry}
Consider the $2\times 2$ matrix $\Sigma$, whose first column is $i \sigma_2 h^\star$, and whose second column is $h$.  In the limit of vanishing gauge couplings, the Standard Model Higgs sector is invariant under
 \begin{equation}
 \Sigma \rightarrow L \Sigma R^\dagger,
 \end{equation}
where $L$ and $R$ and are independent $SU(2)$
global symmetry transformations.
The group $SU(2)_L \times SU(2)_R$ is broken to the diagonal subgroup by
the Higgs VEV $\langle \Sigma \rangle \propto {\bf 1}$, and the unbroken group is called custodial symmetry, or $SU(2)_C$. Under $SU(2)_C$, the three NGBs from $\Sigma$ transform as a triplet.  These become the longitudinal components of the $W^\pm$ and $Z$. 
When the Standard Model gauge couplings are turned on, the Higgs doublet
couples to the $SU(2)_W$ gauge fields through currents $j_\mu^a$, and these
transform as a triplet under $SU(2)_C$ just as the NGBs do. 
The fact that $SU(2)_C$ breaking in the Higgs and gauge boson sectors
arises only from the hypercharge gauge coupling leads to the
tree-level relation $M_W=M_Z \cos \theta$.

Given a modified Higgs sector, there is no guarantee that this relation  will be preserved, unless there is an unbroken custodial $SU(2)$ under which the NGB's and the $SU(2)_W$ currents both transform as triplets.
The heavy $U(1)$ gauge bosons in little Higgs models couple as $T^3_R$, that is, in an $SU(2)_C$-violating fashion, and integrating out these particles out generates dangerous operators such as $(h^\dagger D_\mu h)^2$.  
 Violation of $SU(2)_C$ also may arise from higher-order terms in the non-linear sigma model kinetic term. Although these effects are typically  quite constraining, they can be minimized by building $SU(2)_C$ in as an approximate symmetry of the theory, as was done in Refs.~\cite{Chang:2003un, Chang:2003zn}.  

Ref.~\cite{Chang:2003un} presented a modification of the Minimal Moose in
which the $SU(3)$ global symmetries are replaced by $SO(5)$ global
symmetries, and the $SU(3) \times SU(2) \times U(1)$ gauge group is replaced
by $SO(5) \times SU(2) \times U(1)$.    The $SO(5)$ global symmetry is large
enough to contain $SU(2)_C$, so the $SU(2)_C$-violating contributions from
the non-linear sigma model structure are automatically absent.  The $SU(2)_C$
violating contributions from exchange of the heavy $U(1)$ gauge boson are
partially canceled because this particle is now joined by other
heavy gauge bosons with which it forms a triplet of $SU(2)_C$.
In the limit where the $SO(5)$ gauge coupling becomes large,
these states become nearly degenerate, which suppresses the
$SU(2)_C$-violating effects produced when the triplet is integrated out. 

One complication is that the model has two Higgs doublets, with a potential that requires their VEVs to be misaligned.  This misalignment is its own source of $SU(2)_C$ violation, so that even when the triplet becomes degenerate, integrating it out still yields an $SU(2)_C$-violating term.  In Ref~\cite{Chang:2003zn}, this complication is avoided by constructing a single-Higgs-doublet model with an approximate $SU(2)_C$, an extension of the Littlest Higgs with coset $SO(9)/(SO(5) \times SO(4))$.  

\subsubsection{T-Parity}
\begin{figure}
\vspace{.2in}
\centerline {
\includegraphics[width=4in]{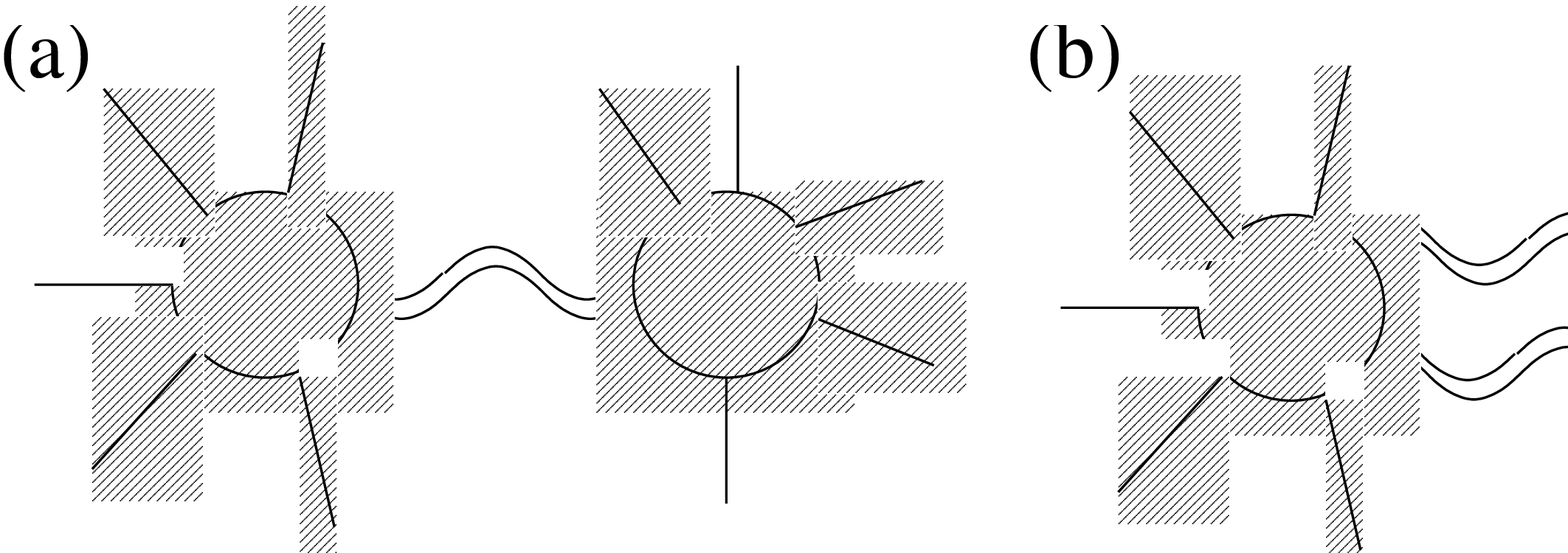}
}
\caption{(a) Tree-level exchange of  T-odd particles is forbidden. (b) Loop processes are allowed.}
\vspace{.2in}
\label{figure:tparity}
\end{figure}

A particularly  interesting class of models
~\cite{Cheng:2003ju, Cheng:2004yc, Low:2004xc} incorporates a discrete symmetry called T-parity, under which the heavy particles are odd, and the Standard Model fields are even.  With this symmetry in place, no effective operators involving just light fields are generated by tree-level exchange of heavy fields, because an even number of heavy fields are required at each vertex (see Fig.~\ref{figure:tparity}).  
As a result, precision electroweak constraints are weakened.   This is similar to the way in which R-parity is helpful for preventing excessive corrections to precision electroweak observables in supersymmetric theories.  

Ref.~\cite{Cheng:2003ju} implemented-T parity in a variation of the Minimal Moose model.  
To avoid $SU(2)_C$-violation from higher-order terms in the non-linear sigma model kinetic term, the global symmetry is enlarged to contain products of $SO(5)$ as in Ref.~\cite{Chang:2003un}, rather than products of $SU(3)$.     If we further modify the Minimal Moose by gauging $[SU(2)\times U(1)]^2$ instead of $SU(3)\times SU(2) \times U(1)$, and set the two $SU(2)$ couplings and the two $U(1)$ couplings equal, then the theory possesses a $Z_2$ symmetry that exchanges the gauge fields of the two $SU(2) \times U(1)$ groups and sends $\Sigma \rightarrow \Sigma^\dagger$, if we neglect the fermions for the time being.  This is apparent  from the reflection symmetry that exchanges the two sites in Fig.~(\ref{fig:moose}).  Denoting one of the gauge fields of the first site by $A_1$ and the analogous gauge field from the second site by $A_2$, the linear combination $A_1+A_2$ is even under this $Z_2$, and the orthogonal combination $A_1-A_2$ is odd.  But $A_1-A_2$ is precisely the linear combination that becomes heavy when the gauge symmetry is broken to the diagonal subgroup: as desired, the heavy gauge boson is odd under T-parity.  

Fermions must be introduced into the picture in a way that preserves the $Z_2$, and so identical copies are introduced on the two sites, {\it i.e.} one copy transforms under one $SU(2) \times U(1)$, and the other transforms under the second $SU(2) \times U(1)$.  To make one linear combination of these fermions heavy, a third site is introduced, with mirror fermions that marry one linear combination.  
Once this is done, it is no longer true that all heavy gauge bosons are T-parity odd, but taking the gauge couplings on the third site to be large makes these T-parity-even gauge bosons extra heavy and suppresses their couplings to Standard Model fermions, which still live on the other two sites.  

Refs.~\cite{Cheng:2004yc, Low:2004xc} explore the potential role of T-parity in little Higgs models further, for instance by elucidating a systematic approach to incorporating fermions in models with T-parity, and by implementing T-parity  in the Littlest Higgs model and in variants of the Littlest Higgs.   
It turns out that it is not possible to implement T-parity in models with a simple gauge group, such as the $SU(3)$ model.

\section{LITTLE HIGGS PHENOMENOLOGY}

\label{sec:pheno}

\subsection{Precision Electroweak Constraints}
As discussed in the introduction and alluded to in the previous section,
indirect constraints from precision electroweak measurements on new
physics at the TeV scale are severe.
The new physics predicted in little Higgs theories is no exception,
and it is important to explore whether there are regions of
parameter space in which these models are consistent with
precision electroweak data.  Of course, agreement with data can
always be achieved by increasing $f$,
and thus the masses of the new particles, but if $f$ becomes too large, this comes at the price of reintroducing fine tuning into the theory\footnote{Although constraints from precision electroweak data are often expressed as lower bounds on $f$, one should keep in mind that the particle masses of the TeV-scale states are a better indicator of how finely tuned the theory is. }.  

There are several sources of corrections to precision electroweak observables in little Higgs theories.  The heavy gauge bosons  couple to fermions  and Higgs doublets through the currents $j_F$ and $j_H$.  After integrating out these gauge bosons one obtains operators of the form $j_F j_F$, $j_F j_H$, and $j_H j_H$.  The $j_F j_F$ terms correspond to new four-fermion operators, which are most strongly constrained by limits from LEPII and from measurements of atomic parity violation, while the $j_F j_H$  terms lead to modifications of the couplings of Standard Model gauge boson  to fermions, and so are constrained by $Z$-pole data.  The $j_H j_H$ terms include the $SU(2)_C$-violating operator $(h^\dagger D_\mu h)^2$, which is generated by heavy $U(1)$ gauge bosons, but not by heavy  $SU(2)$-triplet gauge bosons.

Another potential source of $SU(2)_C$-violation comes from expanding the kinetic term of the non-linear sigma model.
This is not the case for the Littlest Higgs or the Simplest Little Higgs,
but is true, for example, for the Minimal Moose.  If the theory contains triplet scalars that acquire VEVs, as the Littlest Higgs does,  these will also contribute to $SU(2)_C$ violation.   All of the effects discussed so far are tree-level effects that can be analyzed using the type of general effective field theory analysis given in Ref.~\cite{Han:2004az}.  Finally, there are also $SU(2)_C$-violating loop contributions coming from the extended top sector, and depending on the model, from an extended Higgs sector.    

For generic choices of parameters, these effects typically require little Higgs models to have fairly large values of $f$.  For the Littlest Higgs, for instance, it was found in Ref.~\cite{Csaki:2002qg} that constraints from precision electroweak data imply $f>4~{\rm TeV}$, which in turn puts a lower bound on the mass of the heavy quark equal to $5.7~{\rm TeV}$. An even slightly stronger bound was obtained in \cite{Hewett:2002px}, by combining indirect constraints with constraints from direct production of the heavy $U(1)$ gauge boson at the Tevatron. Using the fact that the heavy quark cuts off the quadratic divergence in the top-loop contribution to the Higgs mass-squared, the  bound $m_U > 5.7~{\rm TeV}$ was estimated to correspond to fine tuning at roughly the percent level.     

Simple modifications of the Littlest Higgs  improve the situation without spoiling the stabilization of the electroweak scale.  If we let the light fermions be charged equally under both $U(1)$'s, rather than just under $U(1)_1$, then taking $g'_1= g'_2$ decouples the heavy $U(1)$ gauge boson from $j_F$ and $j_H$ simultaneously, and the precision electroweak problems associated with this particle go away. 
Alternatively, one can just gauge $U(1)_Y$, rather than a product of $U(1)$'s, because the quadratically divergent contribution to the Higgs mass-squared due to hypercharge interactions is numerically small for $\Lambda \sim 10~{\rm TeV}$. In Ref.~\cite{Csaki:2003si} it was shown that if only $U(1)_Y$ is gauged, there is allowed parameter space with $f \sim {\rm TeV}$, with the main constraint coming from the VEV of the triplet scalar.  In the model built on $SU(6)/Sp(6)$, these regions expand  due the absence of the triplet  \cite{Csaki:2003si, Gregoire:2003kr}.

Precision electroweak constraints were also studied for  the Minimal Moose model in Ref.~\cite{Kilic:2003mq}, where it was found that by taking the gauge group to be $[SU(2) \times U(1)]^2$ rather than $SU(3) \times SU(2) \times U(1)$ -- again,  this allows one to choose couplings that effectively decouple the heavy $U(1)$ gauge boson from both $j_F$ and $j_H$ -- it is possible to find regions in parameter space that are consistent with precision electroweak constraints, without severe fine tuning.   Finally, Ref.~\cite{Csaki:2003si} also analyzed contributions to precision electroweak observables in the $SU(3)$ model, and found the constraint $f > 4~{\rm TeV}$; however, in this case, the mass of the heavy fermion may be well below $f$, and so it is possible to have only mild fine tuning even with large values of $f$.   

For more on the indirect effects of new particles in little Higgs theories,  we refer the reader to the literature   \cite{Chivukula:2002ww, Huo:2003vd, Kilic:2003mq, Chen:2003fm, Casalbuoni:2003ft, Kilian:2003xt, Yue:2004xt,  Choudhury:2004ce, Lee:2004me, Buras:2004kq, Choudhury:2004bh, Marandella:2005wd}.  

Although there is parameter space in which the simplest little Higgs theories are consistent with precision electroweak data, it is interesting to consider alternative  models in which the most dangerous effects are automatically absent.  The T-parity conserving models discussed in the previous section are one example. Because tree-level effects associated with the T-odd particles are removed, bounds from precision electroweak data can be satisfied even with $f$ below $1~{\rm TeV}$ \cite{Chang:2003un, Chang:2003zn}.   The models incorporating $SU(2)_C$, also already discussed, are another example.  These provide an alternative to gauging only $U(1)_Y$ by making the heavy $U(1)$ gauge boson a member of an $SU(2)_C$ triplet.

\subsection{Direct Production of Little Higgs Partners}
The spectrum of new particles varies somewhat from one little Higgs model to another, but  all of them predict at least one vector-like quark at the TeV scale, along with extra gauge bosons and scalars.  This is guaranteed given that the little-Higgs mechanism arranges for quadratic divergences to cancel between states of the same statistics.  Here we summarize the prospects for these particles  to be discovered at the LHC.  

Let us focus on theories with a product group structure.  The collider phenomenology of a heavy $U(1)$ gauge boson is certainly  interesting, but since this particle is  associated with the most stringent precision electroweak constraints, and since it is not essential for stabilizing the weak scale, most phenomenological studies have concentrated  on the heavy $SU(2)$ gauge bosons.

At the LHC, these gauge bosons would be produced in $pp$ collisions predominantly through quark-antiquark annihilation.  The production rate depends on $\cot \theta =g_1/g_2$, where $g_1$ is the gauge coupling of the $SU(2)$ under which the Standard Model fermion doublets transform, and $g_2$ is the coupling of the other $SU(2)$;  the fermion couplings to the heavy gauge bosons are proportional to $g \cot \theta$ .   For $\cot \theta=1$, the cross section for producing a 3 TeV neutral $Z_H$ at the LHC is $\sim 100~{\rm fb }$ \cite{Burdman:2002ns, Han:2003wu}, corresponding to tens of thousands of events.  A clean discovery channel would arise from the decay of $Z_H$ to pairs of highly energetic leptons, and the discovery reach for $\cot \theta=1$ is roughly $5~\rm{TeV}$.  Precision electroweak data prefer smaller values, $\cot\theta \lsim .2$, 
but even for these values, the discovery reach is still well into the multi-TeV region.  After ameliorating or eliminating the precision electroweak constraints associated with a heavy $U(1)$ gauge boson, the lower bound on the mass of the heavy $SU(2)$ gauge bosons is roughly $2~{\rm TeV}$, leaving plenty of room for discovery.  

The phenomenology of the heavy neutral gauge boson of the $SU(3)$ model
(Simplest Little Higgs) is similar to the phenomenology of the $Z_H$ in
product gauge group models.  On the other hand, the detection of the
heavy $SU(2)$-doublet gauge bosons of the $SU(3)$ model is complicated
by a $v/f$ suppression of the coupling of these states to light quarks.  

Even if heavy gauge bosons are discovered, there is still the question of whether they arise from a little Higgs model.  
Studying the Littlest Higgs, the authors of Ref.~\cite{Burdman:2002ns} pointed out that the partial width of the $Z_H$ to fermions
is proportional to  $\cot^2 \theta$, while the partial width into boson pairs $Z h$ and $W^+ W^-$  is proportional to $\cot^2 2\theta$. 
This feature was proposed as an interesting test of the little Higgs structure, because
in an alternative theory with the Higgs  charged under just  one $SU(2)$, the partial width into bosons is proportional to $\cot^2 \theta$ just like for the fermions.

The production of the heavy quark $(U,U^c)$ in the Littlest Higgs model was also studied in Refs.~\cite{Han:2003wu, Perelstein:2003wd}. These  can be pair-produced  via their coupling to gluons, but due to the mixing in the top quark sector, they  can also be produced singly via $Wb$ fusion, $W^+b \rightarrow U$ \cite{Han:2003wu}.  This mode, whose rate depends on the ratio of the $\lambda$ and $\lambda'$ couplings that appear in the top Yukawa sector of the model, tends to dominate for larger values of the mass, $m_U \gsim 1~{\rm TeV}$.   The heavy quark branching fractions satisfy $\Gamma(U \rightarrow t h) \approx \Gamma(U \rightarrow t Z) \approx {1\over 2} \Gamma(U \rightarrow b W^+)$, and all three decay modes lead to identifiable signatures.
The discovery reach can be estimated to be roughly $3~{\rm TeV}$.

The parameters $\lambda$ and $\lambda'$ determine the top Yukawa coupling,  and along with $f$, they also determine the mass of $U$ and its coupling to light states.  The decay constant $f$ can be determined by  measurements of the properties of the heavy gauge bosons \cite{Burdman:2002ns}, and the known top Yukawa coupling gives one equation constraining $\lambda$ and $\lambda'$.  By measuring $m_U$ one would obtain a second, independent equation.  Ref.~\cite{Perelstein:2003wd} considered whether it might be possible to test the structure of the  Littlest Higgs top sector by combining the result of the measurement of $m_U$ with a  measurement of  the $U$ production cross section, which depends on  $\lambda$ and $\lambda'$ in yet another way.   This  would be another  interesting experimental probe of the little Higgs mechanism for canceling quadratic divergences, this time in the top sector.
The authors concluded that  such a test might be feasible if the uncertainty in the bottom-quark parton distribution function were reduced.

The prospects for discovering  extra scalar particles are quite model dependent.  For the Littlest Higgs, Ref.~\cite{Han:2003wu} pointed out the possibility that the doubly charged scalar belonging to the triplet $\phi$ would mediate a resonant contribution to longitudinal $WW$ scattering, possibly giving a signal above background.  In other little Higgs theories, for example $SU(6)/Sp(6)$, the triplet is replaced by a neutral scalar, whose discovery prospects look grim (that model has two Higgs doublets, so the Higgs phenomenology would still be different than in the Standard Model).

Finally, we should mention that the phenomenology of little Higgs theories is drastically altered if the theory conserves T-parity, in which case there are missing-energy signals that are similar to those in supersymmetric theories with R-parity conservation \cite{Hubisz:2004ft} . Other work on little Higgs phenomenology appears in Refs.~\cite{Han:2003gf, Dib:2003zj, Sullivan:2003xy, Yue:2003yk, Azuelos:2004dm, Logan:2004hj, Gonzalez-Sprinberg:2004bb, Cho:2004ap, Kilian:2004pp, Yue:2004fv, Park:2004ab}.

\section{CONCLUSIONS}
Little Higgs theories are a compelling possibility for physics beyond the Standard Model.
These theories feature weakly coupled new physics at the TeV scale, which  stabilizes the Higgs potential even with a cut-off as large as $\sim 10~\rm{TeV}$.  The key ingredient is that the Higgs is a pseudo-Nambu-Goldstone boson of a spontaneously broken approximate global symmetry, with the explicit breaking of this symmetry collective in nature -- that is, more than one coupling at a time must be turned on for the  symmetry to be broken. The collective breaking ensures that no quadratically divergent contributions to the Higgs potential arise at one loop.   New TeV-scale particles cancel the quadratic divergences of Standard Model fields with the same statistics, and some of these new particles should be revealed at the LHC if they play a role in stabilizing the weak scale.  

For little Higgs theories with T-parity or any other unbroken discrete symmetry at the
TeV scale, the lightest particle charged under the symmetry is stable,
and might play an important role in cosmology. If it is electrically
neutral it may well be a good cold dark matter candidate \cite{Birkedal-Hansen:2003mp}.

Little Higgs theories are effective field theories valid up to a cut-off $\Lambda \sim 4 \pi f$.  An important question that this review has not addressed is what lies beyond the cut-off $\Lambda$.  This question has not been explored extensively in the literature, but we will conclude here by mentioning a few possibilities.  One possibility is that that the global symmetry is broken by a weakly coupled scalar, and this scalar's potential is protected by its own little Higgs mechanism  -- it is a pseudo-NGB of a {\it different} symmetry.  By building a structure with a single iteration of this type one can raise the cut-off to $\Lambda \sim 100~\rm{TeV}$ \cite{Kaplan:2004cr}, and more ambitiously, one can attempt to construct a theory with many iterations, with a much higher cut-off \cite{Batra:2004ah}.   A different possibility is that the global symmetry is broken by strong dynamics that give rise to  fermion condensation.  In this case the Higgs is a composite particle.  An explicit little Higgs UV completion of this type was constructed in Ref.~\cite{Katz:2003sn}, which employs soft supersymmetry breaking at $\sim 10~{\rm TeV}$ to generate the four-fermion operators required to give masses to the Standard Model fermions.   Finally, theories have been constructed in five-dimensional Anti-de Sitter space that can be interpreted as holographic duals of composite Higgs theories \cite{Contino:2003ve,Agashe:2004rs,Thaler:2005en}, and these theories can also be thought of as UV completions that involve strong dynamics at the scale $\Lambda$.  


\end{document}